\documentclass[longauth]{aa}
\usepackage{graphicx}
\usepackage{txfonts}
\usepackage{xcolor}
\usepackage{hyperref}
\hypersetup{colorlinks, linkcolor=blue, citecolor=blue, urlcolor=blue}
\usepackage{amsmath}
\usepackage{ulem}
\usepackage{soul}
\usepackage{enumitem}

\newcommand{\pivec}{\mbox{\boldmath $\pi$}}
\newcommand{\muvec}{\mbox{\boldmath $\mu$}}

\newcommand{\te}{t_{\rm E}}
\newcommand{\thetae}{\theta_{\rm E}}

\newcommand{\pie}{\pi_{\rm E}}

\newcommand{\dl}{D_{\rm L}}
\newcommand{\ds}{D_{\rm S}}

\definecolor{brown}{rgb}{0.59, 0.29, 0.0}
\definecolor{darkgreen}{rgb}{0.0, 0.42, 0.24}
\definecolor{darkblue}{rgb}{0.01, 0.31, 0.59}
\definecolor{darkblue}{rgb}{0.0, 0.25, 0.42}
\definecolor{blue}{rgb}{0.0,0.0,1.0}
\definecolor{green}{rgb}{0.0,1.0,0.0}

\begin{document}

\title{KMT-2023-BLG-0416, KMT-2023-BLG-1454, KMT-2023-BLG-1642: 
Microlensing planets identified from partially covered signals}
\titlerunning{KMT-2023-BLG-0416, KMT-2023-BLG-1454, and KMT-2023-BLG-1642}

\author{
     Cheongho~Han\inst{\ref{inst1}} 
\and Andrzej~Udalski\inst{\ref{inst2}} 
\and Chung-Uk~Lee\inst{\ref{inst3}} 
\and Weicheng~Zang\inst{\ref{inst4}}     
\\
(Leading authors)\\
     Michael~D.~Albrow\inst{\ref{inst5}}   
\and Sun-Ju~Chung\inst{\ref{inst3},\ref{inst6}}      
\and Andrew~Gould\inst{\ref{inst7},\ref{inst8}}      
\and Kyu-Ha~Hwang\inst{\ref{inst3}} 
\and Youn~Kil~Jung\inst{\ref{inst3}} 
\and Yoon-Hyun~Ryu\inst{\ref{inst3}} 
\and Yossi~Shvartzvald\inst{\ref{inst9}}   
\and In-Gu~Shin\inst{\ref{inst10}} 
\and Jennifer~C.~Yee\inst{\ref{inst10}}   
\and Hongjing~Yang\inst{\ref{inst4}}     
\and Sang-Mok~Cha\inst{\ref{inst3},\ref{inst11}} 
\and Doeon~Kim\inst{\ref{inst1}}
\and Dong-Jin~Kim\inst{\ref{inst3}} 
\and Seung-Lee~Kim\inst{\ref{inst3},\ref{inst6}} 
\and Dong-Joo~Lee\inst{\ref{inst3}} 
\and Yongseok~Lee\inst{\ref{inst3},\ref{inst11}} 
\and Byeong-Gon~Park\inst{\ref{inst3},\ref{inst6}} 
\and Richard~W.~Pogge\inst{\ref{inst8}}
\\
(The KMTNet Collaboration),\\
     Przemek~Mr{\'o}z\inst{\ref{inst2}} 
\and Micha{\l}~K.~Szyma{\'n}ski\inst{\ref{inst2}}
\and Jan~Skowron\inst{\ref{inst2}}
\and Rados{\l}aw~Poleski\inst{\ref{inst2}} 
\and Igor~Soszy{\'n}ski\inst{\ref{inst2}}
\and Pawe{\l}~Pietrukowicz\inst{\ref{inst2}}
\and Szymon~Koz{\l}owski\inst{\ref{inst2}} 
\and Krzysztof~A.~Rybicki\inst{\ref{inst2},\ref{inst9}}
\and Patryk~Iwanek\inst{\ref{inst2}}
\and Krzysztof~Ulaczyk\inst{\ref{inst12}}
\and Marcin~Wrona\inst{\ref{inst2}}
\and Mariusz~Gromadzki\inst{\ref{inst2}}          
\and Mateusz Mr{\'o}z\inst{\ref{inst2}}
\\
(The OGLE Collaboration)\\
}

\institute{
     Department of Physics, Chungbuk National University, Cheongju 28644, Republic of Korea                              \label{inst1}      
\and Astronomical Observatory, University of Warsaw, Al.~Ujazdowskie 4, 00-478 Warszawa, Poland                          \label{inst2}      
\and Korea Astronomy and Space Science Institute, Daejon 34055, Republic of Korea                                        \label{inst3}      
\and Department of Astronomy and Tsinghua Centre for Astrophysics, Tsinghua University, Beijing 100084, China            \label{inst4}      
\and University of Canterbury, Department of Physics and Astronomy, Private Bag 4800, Christchurch 8020, New Zealand     \label{inst5}      
\and Korea University of Science and Technology, 217 Gajeong-ro, Yuseong-gu, Daejeon, 34113, Republic of Korea           \label{inst6}      
\and Max Planck Institute for Astronomy, K\"onigstuhl 17, D-69117 Heidelberg, Germany                                    \label{inst7}      
\and Department of Astronomy, The Ohio State University, 140 W. 18th Ave., Columbus, OH 43210, USA                       \label{inst8}      
\and Department of Particle Physics and Astrophysics, Weizmann Institute of Science, Rehovot 76100, Israel               \label{inst9}      
\and Center for Astrophysics $|$ Harvard \& Smithsonian 60 Garden St., Cambridge, MA 02138, USA                          \label{inst10}     
\and School of Space Research, Kyung Hee University, Yongin, Kyeonggi 17104, Republic of Korea                           \label{inst11}     
\and Department of Physics, University of Warwick, Gibbet Hill Road, Coventry, CV4 7AL, UK                               \label{inst12}     
}
\date{Received ; accepted}

\abstract
{}
{
We investigate the 2023 season data from high-cadence microlensing surveys with the aim of
detecting partially covered short-term signals and revealing their underlying astrophysical
origins. Through this analysis, we ascertain that the signals observed in the lensing events
KMT-2023-BLG-0416, KMT-2023-BLG-1454, and KMT-2023-BLG-1642 are of planetary origin.
}
{
Considering the potential degeneracy caused by the partial coverage of signals, we thoroughly 
investigate the lensing-parameter plane.  In the case of KMT-2023-BLG-0416, we have identified 
two solution sets, one with a planet-to-host mass ratio of $q\sim 10^{-2}$ and the other with 
$q\sim 6\times 10^{-5}$, within each of which there are two local solutions emerging due to the 
inner-outer degeneracy.  For KMT-2023-BLG-1454, we discern four local solutions featuring mass 
ratios of $q\sim (1.7-4.3)\times 10^{-3}$.  When it comes to KMT-2023-BLG-1642, we identified 
two locals with $q\sim (6-10)\times 10^{-3}$ resulting from the inner-outer degeneracy. 
}
{
We estimate the physical lens parameters by conducting Bayesian analyses based on the event 
time scale and Einstein radius.  For KMT-2023-BLG-0416L, the host mass is $\sim 0.6~M_\odot$, 
and the planet mass is $\sim (6.1-6.7)~M_{\rm J}$ according to one set of solutions and 
$\sim 0.04~M_{\rm J}$ according to the other set of solutions.  KMT-2023-BLG-1454Lb has a mass 
roughly half that of Jupiter, while KMT-2023-BLG-1646Lb has a mass in the range of between 1.1 
to 1.3 times that of Jupiter, classifying them both as giant planets orbiting mid M-dwarf host 
stars with masses ranging from 0.13 to 0.17 solar masses.
}
{}


\maketitle

\section{Introduction}\label{sec:one}

The microlensing signature of a planet, in general, manifests as a brief transient anomaly within
the lensing light curve produced by the host of the planet \citep{Mao1991, Gould1992b}. Detecting 
such a brief signal posed a formidable challenge in the early stage of microlensing experiments, 
such as the Optical Gravitational Lensing Experiment \citep[OGLE:][]{Udalski1994} and Massive 
Astrophysical Compact Halo Object \citep[MACHO:][]{Alcock1995} experiments, in which lensing 
events were observed at approximately one-day intervals.  In the 1990s and 2000s, the necessary 
observational frequency to detect planetary signals was achieved through the implementation of early 
warning systems, as exemplified by the pioneering work of \citet{Udalski1994} and \citet{Alcock1996}.  
These systems were coupled with subsequent observations of alerted events, conducted by various 
follow-up teams, including Galactic Microlensing Alerts Network \citep[GMAN:][]{Alcock1997}, Probing 
Lensing Anomalies NETwork \citep[PLANET:][]{Albrow1998}, Microlensing Follow-Up Network 
\citep[$\mu$FUN:][]{Yoo2004b}, and RoboNet \citep{Tsapras2003}. For an in-depth exploration on 
the historical development of the observational strategies implemented during this era, we recommend 
referring to the comprehensive review of \citet{Gould2010}.

In contemporary microlensing experiments, the essential high-frequency observations necessary for
capturing short planetary signals are made possible through utilizing a network of globally dispersed 
telescopes which are equipped with very wide-field cameras. Regarding the Korea Microlensing Telescope 
Network \citep[KMTNet:][]{Kim2016} survey, the observational frequency in its primary fields is reduced 
to as little as 15 minutes, a time frame sufficiently brief to detect signals generated by a planet 
with a mass similar to that of Earth. With the capability to continuously monitor all lensing events 
and discern signals emerging from various segments of lensing light curves, contemporary lensing 
surveys are now detecting an estimated 30 planets on an annual basis \citep{Gould2022a, Gould2022b, 
Jung2022}.

Although the current microlensing surveys have improved their observational cadence, there are
still gaps in the coverage of a portion of planetary signals. Weather conditions at telescope 
sites are the primary factor responsible for the incomplete coverage of planetary signals. 
Planetary signals typically exhibit durations spanning several days for giant planets and several 
hours for terrestrial planets, and incomplete coverage of these signals occasionally occurs when 
adverse weather conditions at telescope sites across the global network hinder observations.  The 
discontinuous coverage of planetary signals can also be attributed to another factor: the temporal 
gaps between observation times at different telescope sites. In practical terms, this means that 
some lensing events can be observable for only a portion of a night, resulting in a time lag between 
observations at one site and the subsequent observations at another site. The significance of this 
gap in observation times is particularly pronounced during the initial and concluding phases of the 
bulge season when the available observation periods are relatively short. Finally, planetary signals 
from events occurring in certain fields covered with relatively low cadences tend to be incomplete, 
particularly for events characterized by very brief anomalies in duration. It is crucial to scrutinize 
partially covered signals because planets associated with these signals might remain unreported 
without in-depth analyses. Neglecting such analyses could lead to inaccurate estimations of the 
detection efficiency, which is a fundamental element for establishing the demographics of planets 
detected through microlensing.

In this paper, we report the discoveries of three microlensing planets from the analyses of the 
lensing events KMT-2023-BLG-0416, KMT-2023-BLG-1454, and KMT-2023-BLG-1642, for which the planetary 
signals within the lensing light curves were only partially observed. The event analyses were 
carried out as a component of a project focused on exploring brief transient anomalies with 
limited data coverage, aiming to identify potential planetary candidates. The initial publication 
of planetary discoveries stemming from the analysis of data gathered during the 2021 and 2022 
seasons of the KMTNet survey was documented in \citet{Han2023a}. The analyses presented in this 
work constitute the project's second release of results, this time stemming from the investigation 
of data gathered during the 2023 season of the KMTNet survey.

\begin{figure}[t]
\includegraphics[width=\columnwidth]{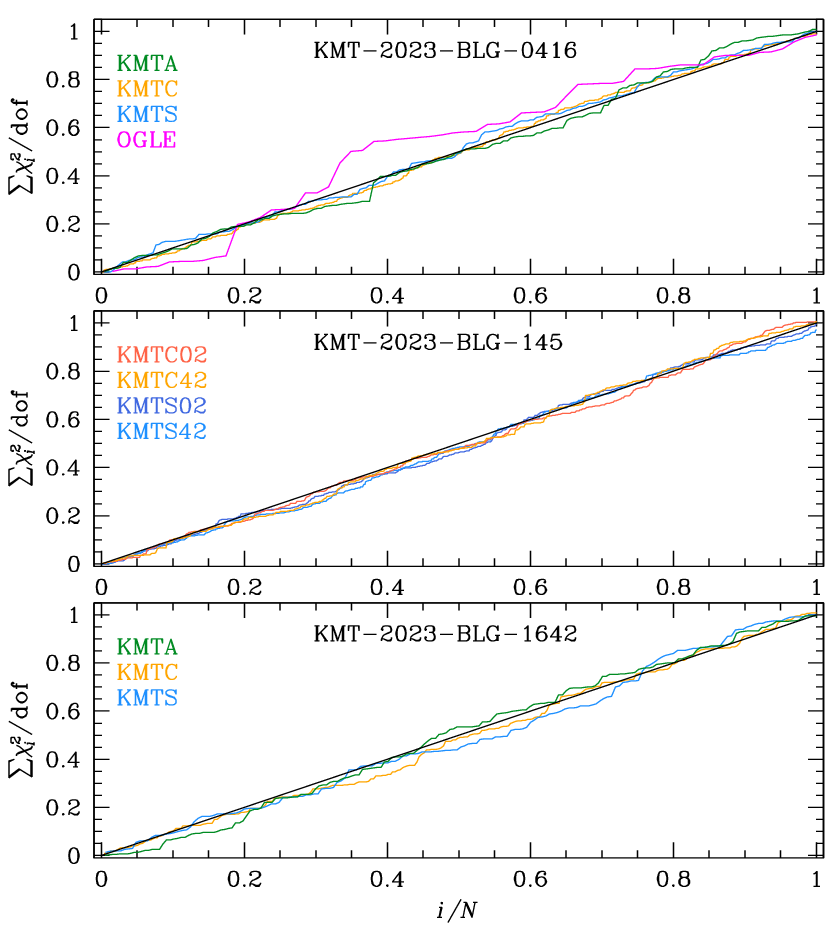}
\caption{
Cumulative sum of $\chi^2$ for the individual datasets of the three analyzed events KMT-2023-BLG-0416, 
KMT-2023-BLG-1454, and KMT-2023-BLG-1642.  The colors of the distributions match those of the 
datasets marked in the legend.
}
\label{fig:one}
\end{figure}

\section{Data and observations}\label{sec:two}

The anomalies in the three microlensing events KMT-2023-BLG-0416, KMT-2023-BLG-1454, and 
KMT-2023-BLG-1642 were found through the examination of lensing events detected during the 2023 
season by the KMTNet survey. This survey focuses on the monitoring of stars lying toward the Galactic 
bulge field in the quest for gravitational lensing events. To achieve continuous coverage of lensing 
events, the survey employs three identical telescopes positioned across the three continents in the 
Southern Hemisphere. The sites of the individual KMTNet telescopes are the Cerro Tololo Interamerican 
Observatory in Chile (KMTC), the South African Astronomical Observatory in South Africa (KMTS), and 
the Siding Spring Observatory in Australia (KMTA).  Each of the KMTNet telescopes has a 1.6 m aperture 
and is equipped with a wide-field camera that delivers a field of view spanning 4 square degrees.  
The survey primarily employs the $I$ band for image acquisition, with approximately ten percent of 
the images obtained in the $V$ band, specifically for the purpose of source color measurement. The 
observation cadences vary depending on the events: 1.0~hour for KMT-2023-BLG-0416, 0.25~hour for 
KMT-2023-BLG-1454, and 2.5~hour for KMT-2023-BLG-1642.  Among the events, KMT-2023-BLG-0416 was 
additionally identified by the OGLE group.  During its fourth phase, the OGLE survey is conducted 
with the use of the 1.3-meter telescope that lies at the Las Campanas Observatory in Chile.  The 
OGLE telescope is equipped with a camera that provides a field of view encompassing 1.4 square 
degrees.  In the analysis of KMT-2023-BLG-0416, we include the OGLE data.

The preliminary image reduction and source photometry were conducted using the data processing
pipelines developed by \citet{Albrow2009} for the KMTNet survey and \citet{Udalski2003} for the
OGLE survey.  For the use of optimal data in the analyses, the KMTNet data were reprocessed using 
the updated KMTNet pipeline, which was recently developed by \citet{Yang2023} and planned to be 
implemented from the 2024 season.  To construct color-magnitude diagrams (CMDs) for stars  lying 
near the source stars and to derive source magnitudes in the $I$ and $V$ passbands, supplementary 
photometry was carried out on the KMTC dataset using the pyDIA code developed by \citet{Albrow2017}. 
We re-calibrated the error bars associated with the photometry data obtained from the pipelines in 
accordance with the procedure outlined in \citet{Yee2012}.  In this procedure, the error bars are 
normalized by $\sigma_i^\prime = k(\sigma_i^2 + \sigma_{\rm min}^2)^{1/2}$, where $\sigma_i$ 
represents the photometry error estimated from the pipeline.  The values of $k$ and $\sigma_{\rm min}$ 
are chosen so that the $\chi^2$ value per degree of freedom (dof) for each dataset becomes unity and 
the cumulative sum of $\chi^2$ is approximately linear as a function of source magnification.  The 
cumulative sum of $\chi^2$ is constructed by first sorting the data points of each dataset by 
magnification, calculating the value of $\chi_i^2$ contributed by each point, and then plotting 
$\sum_i^N \chi_i^2$ as a function of $N$.  Here $N$ denotes the number of data points with 
magnification less than or equal to the magnification of point $N$.  Figure~\ref{fig:one} shows 
the cumulative sum of $\chi^2$ for the individual datasets of the three analyzed events.  This 
calibration was performed to ensure that the error bars are consistent with the scatter of data.

\section{Light curve analyses}\label{sec:three}

It turns out that the anomalies in the light curves of the analyzed lensing events are very 
likely to be generated by planetary companions to the lenses.  We start by offering an 
introductory overview of the fundamental physics of planetary lensing, which serves to introduce 
pertinent terminology, elucidate the modeling procedure, and clarify the parameters employed 
in the modeling process, before proceeding with in-depth event analyses.

A planet-induced anomaly arises when the source of a lensing event passes over or approaches
close to the caustics induced by the planet. Caustics represent positions on the source plane
where the lensing magnification of a point source becomes infinitely large.  In the case of a 
lens composed of a single planet and its host, with a planet/host mass ratio less than order 
of $10^{-3}$, the lensing behavior is described by a binary-lens single-source (2L1S) formalism 
with a very low mass ratio.  Typically, a planet generates two distinct sets of caustics: one 
set, known as the central caustic, is situated in close proximity to the position of the planet 
host, while the other set, referred to as the planetary caustic, is located away from the host 
at a position approximately $\sim (1-1/s^2) {\bf s}$.\footnote{ In the case where the planet-host 
separation is very close to the Einstein ring, the central and planetary caustics merge to form 
a single caustic curve.  } Here ${\bf s}$ denotes the position vector of the planet from the 
host with its length normalized to the angular Einstein radius $\thetae$ of the lens system. 
From the perspective of the lens plane, planetary anomalies occur when the planet perturbs one 
of the two source images formed by the planet's host. When the planet perturbs the brighter 
(major) image, the image is further magnified and the resulting anomaly exhibits a positive 
deviation. On the other hand, when the planet perturbs the fainter (minor) image, the image 
is often demagnified, which then results in a negative deviation \citep{Gaudi1997}.

The number and shape of the planet-induced caustics vary depending on the projected separation
and mass ratio between the planet and host. Planet-induced caustics are classified into three 
topologies: "close", "wide", and "resonant" \citep{Erdl1993}.  A close planet induces a pair of 
planetary caustics positioned on the opposite side of the planet with respect to the primary, 
whereas a wide planet induces a single planetary caustic on the planet side.  In the case of a 
resonant caustic, the planetary and central caustics merge together, resulting in a unified set 
of caustics. The detailed descriptions on the shapes and characteristics of the central and 
planetary caustics are extensively covered in the works of \citet{Chung2005} and \citet{Han2006}, 
respectively.

Assuming that the relative motion between the lens and source is rectilinear, the light curve 
of a planetary event can be described using seven parameters. The first set of three parameters 
$(t_0, u_0, \te)$ characterizes the approach of the source to the lens, and the individual 
parameters denote the time of the closest approach, the lens-source separation at $t_0$ (impact 
parameter, scaled to $\thetae$), and the event time scale. Another set of three parameters $(s, q, 
\alpha)$ characterize the binary lens, and the parameters denote the projected separation 
(normalized to $\thetae$), the mass ratio, and the angle between the source trajectory and the 
binary-lens axis (source trajectory angle), respectively.  The last parameter $\rho$, which is 
defined as the ratio of the angular source radius $\theta_*$ to $\thetae$, serves to describe the 
deformation of the lensing light curve by finite-source effects, which become prominent as the 
source approaches very near to the caustic or passes over it.

\begin{figure}[t]
\includegraphics[width=\columnwidth]{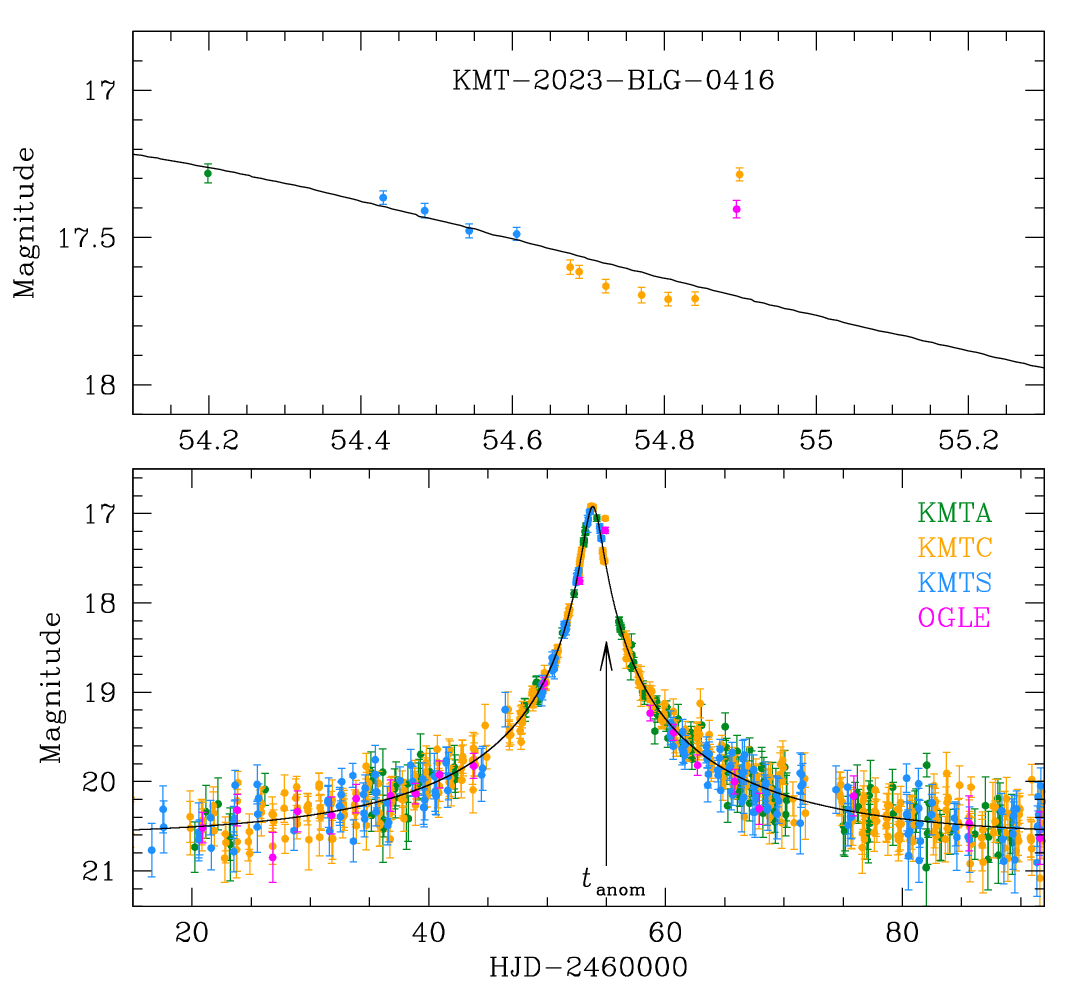}
\caption{
Lensing light curve of KMT-2023-BLG-0416. The lower panel presents the overall perspective, 
while the upper panel provides a close-up view of the anomaly region. The arrow in the lower 
panel indicates the time of the anomaly. The curve drawn over the data points is a 1L1S model 
obtained by fitting the light curve with the exclusion of the data around the anomaly. 
}
\label{fig:two}
\end{figure}

We analyze the events by modeling the light curves in search for the lensing solutions of the 
individual events. The term "lensing solution" refers to the collection of lensing parameters 
that most accurately describe the observed light curve. In the 2L1S modeling, the searches for 
the parameters are done in two steps. In the first step, we seek the binary parameters $s$ and 
$q$ through a grid-based approach, employing multiple initial values of $\alpha$. Subsequently, 
we determine the remaining parameters through a downhill method. In this downhill approach, we 
find the lensing parameters by minimizing $\chi^2$ using the Markov Chain Monte Carlo (MCMC) 
technique, which incorporates a Gaussian sampler with an adaptable step size, as outlined in 
\citet{Doran2004}. The initial step in this procedure results in a $\Delta\chi^2$ map across the 
grid parameters, that is, $(s, q)$. We identify local solutions that manifest on the $(s, q)$ 
parameter plane.  In the second step, we further refine the initially identified local solutions 
and determine a global solution by evaluating the goodness of fit across these local solutions. 
In the case of the lensing events examined in this study, the observed anomalies on the light 
curves are only partially observable, potentially leading to various forms of degeneracy when 
interpreting these anomalies.  Consequently, it becomes crucial to investigate and consider 
these degenerate solutions in order to make accurate interpretations of the observed anomalies.

\begin{table*}[t]
\footnotesize
\caption{Solutions of KMT-2023-BLG-0416}\label{table:one}
\begin{tabular}{lllll}
\hline\hline
\multicolumn{1}{c}{Parameter}                    &
\multicolumn{1}{c}{Local A$_{\rm in}$}           &
\multicolumn{1}{c}{Local A$_{\rm out}$}          &
\multicolumn{1}{c}{Local B$_{\rm in}$}           &
\multicolumn{1}{c}{Local B$_{\rm out}$}          \\
\hline
$\chi^2/{\rm dof}$            &  1379.3/1368           &  1372.0/1368          &    1366.4/1368           &  1365.9/1368            \\
$t_0$ (HJD$^\prime$)          &  $54.154 \pm  0.041$   &  $54.183 \pm 0.031 $  &    $53.859 \pm  0.007$   &  $53.858 \pm 0.007$     \\
$u_0$ ($10^{-2}$)             &  $ 2.26  \pm  0.23 $   &  $ 2.36  \pm 0.16  $  &    $ 3.25  \pm  1.82 $   &  $ 3.34  \pm 0.19 $     \\
$\te$ (days)                  &  $ 29.91 \pm  2.17 $   &  $27.15  \pm 1.57  $  &    $24.33  \pm  1.23 $   &  $23.72  \pm 1.23 $     \\
$s$                           &  $ 1.368 \pm  0.040$   &  $ 0.774 \pm 0.017 $  &    $ 1.054 \pm  0.005$   &  $ 0.997 \pm 0.005$     \\
$q$ ($10^{-3}$)               &  $ 9.66  \pm  1.55 $   &  $ 9.98  \pm 1.06  $  &    $ 0.066 \pm  0.010$   &  $ 0.062 \pm 0.008$     \\
$\alpha$ (rad)                &  $ 3.657 \pm  0.019$   &  $ 3.628 \pm 0.021 $  &    $ 3.786 \pm  0.006$   &  $ 3.793 \pm 0.007$     \\
$\rho$ ($10^{-3}$)            &  $ 1.75  \pm  0.35 $   &  $ 1.41  \pm 0.28  $  &    $ 1.58  \pm  0.22 $   &  $ 1.36  \pm 0.20 $     \\
\hline                                                 
\end{tabular}
\tablefoot{ ${\rm HJD}^\prime = {\rm HJD}- 2460000$.  }
\end{table*}

\subsection{KMT-2023-BLG-0416}\label{sec:three-one}

The source star of the lensing event KMT-2023-BLG-0416 lies at the equatorial coordinates 
$({\rm RA}, {\rm DEC}) = $ (17:45:17.77, $-25$:04:46.16), which correspond to the Galactic 
coordinates $(l, b) = (3^\circ\hskip-2pt .255, 2^\circ\hskip-2pt .071)$.  The $I$-band 
magnitude of the source before the lensing magnification, baseline magnitude, is $I_{\rm base}
=20.41$, and the extinction toward the field is $A_I=3.02$. The event was first found by the 
KMTNet group on 2023 April 17, which corresponds to the abridged heliocentric Julian date 
${\rm HJD}^\prime\equiv {\rm HJD}-2460000\sim 51$. At the time of finding the event, the 
source was brighter than the baseline magnitude by $\Delta I\sim 1.86$. The OGLE group 
independently detected the event 10 days after its initial discovery by KMTNet and labeled 
the event as OGLE-2023-BLG-0402.  Hereafter we refer to the event using the event ID provided 
by the KMTNet group in accordance with the convention that adopts the event ID assigned by 
the initial discovery group.

\begin{figure}[t]
\includegraphics[width=\columnwidth]{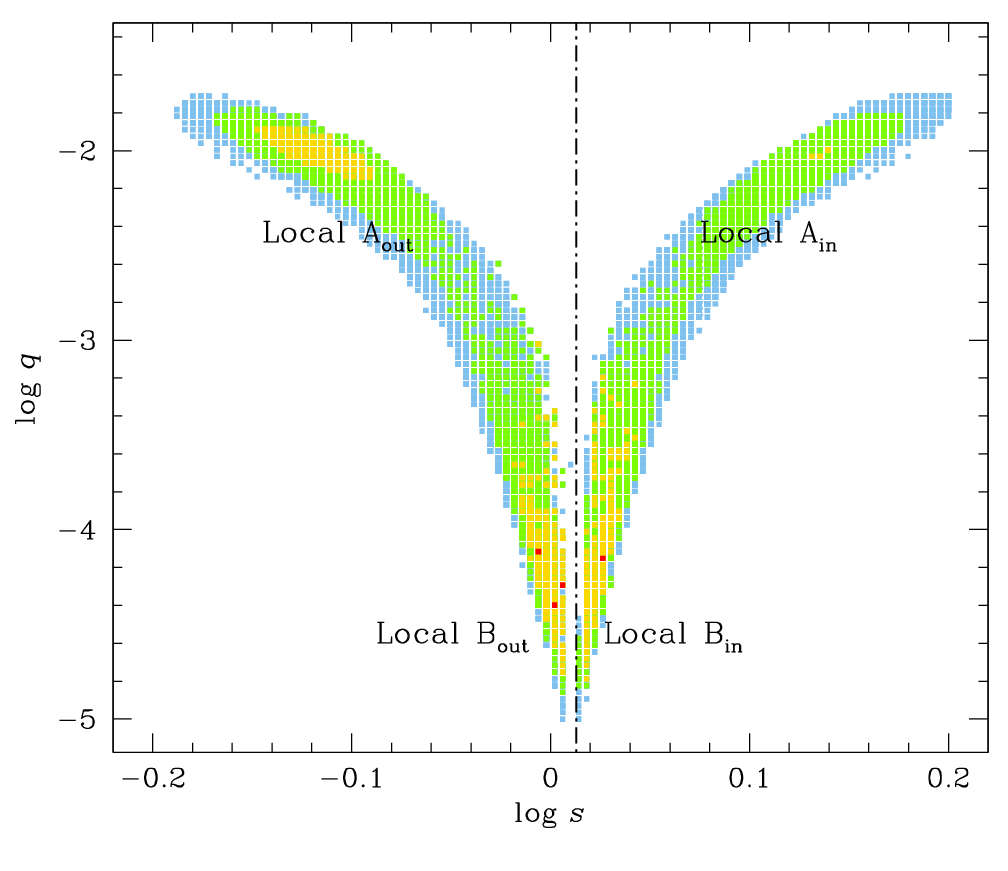}
\caption{
$\Delta\chi^2$ map on the plane of the planet parameters $(\log s, \log q)$ of the lensing 
event KMT-2023-BLG-0416. The color scheme represents points with $\Delta\chi^2 < 1^2n$ (red), 
$< 2^2n$ (yellow), $< 3^2n$ (green), and $< 4^2n$ (cyan), where $n=2$.  The labels marked by 
A$_{\rm in}$, A$_{\rm out}$, B$_{\rm in}$, and B$_{\rm out}$ represent positions of the four 
local solutions. The vertical dot-dashed line indicates the geometric mean between the binary 
separations of the B$_{\rm in}$ and B$_{\rm out}$ solutions. 
}
\label{fig:three}
\end{figure}

The lensing light curve of KMT-2023-BLG-0416 constructed with the combined KMTNet and OGLE 
data is shown in Figure~\ref{fig:two}. The lower panel provides an overview, while the upper 
panel offers a close-up view of the area indicated by the arrow in the lower panel.  The 
curve drawn over the data points is the single-lens single-source (1L1S) model obtained by 
fitting the data excluding those around the anomaly. From the detailed inspection of the 
light curve, we found that the light curve exhibited a partially covered short-term anomaly 
in the region around ${\rm HJD}^\prime \sim 54.9$. The ascending segment of the anomaly was 
captured by the data obtained through observations carried out with the Chilean telescopes, 
specifically KMTC and OGLE data. However, the subsequent portion of the anomaly remained 
unobserved due to cloud cover at the Australian observation site. Although only a small portion 
of the anomaly was covered, the signal is very likely to be real because both the KMTC and OGLE 
observations verified the ascending segment of the anomaly. In addition to the ascending 
segment, the anomaly displays minor negative deviations preceding the rise, and this suggests
that the rising anomaly pattern resulted from the source crossing over a caustic induced by 
a lens companion.

\begin{figure}[t]
\includegraphics[width=\columnwidth]{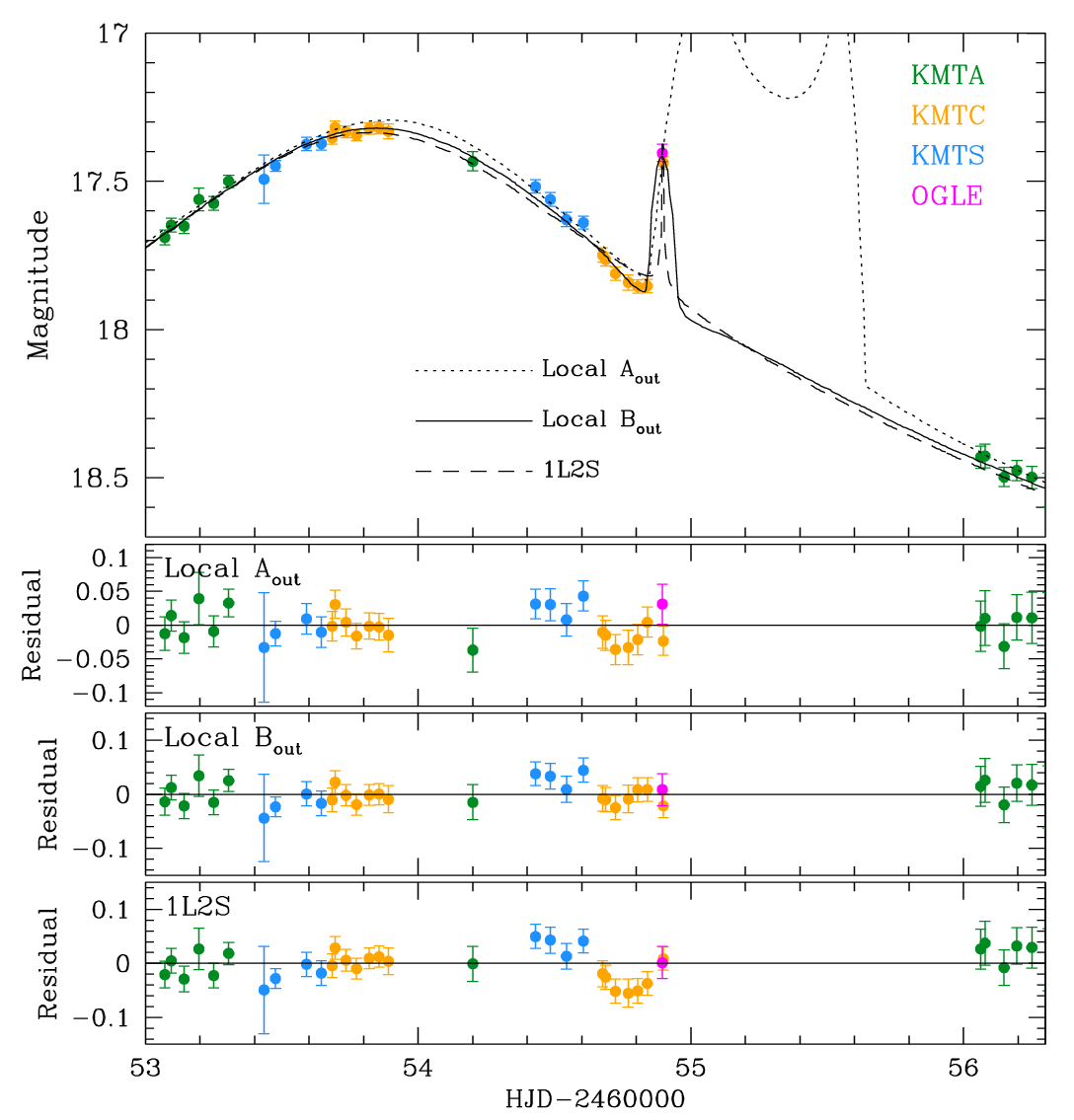}
\caption{
Model curves of the local solutions A$_{\rm out}$ and B$_{\rm out}$ for KMT-2023-BLG-0416. 
The lower panels show the residuals from the two local planetary solutions and the 
single-lens binary-source (1L2S) solution.
}
\label{fig:four}
\end{figure}

Considering the caustic-related feature of the anomaly, we conducted a 2L1S modeling of the 
event.  For the thorough exploration of all potential solutions capable of explaining the 
observed anomaly, we conducted an extensive grid search for the binary parameters $s$ and $q$. 
Figure~\ref{fig:three} shows the $\Delta\chi^2$ map on the $(\log s, \log q)$ parameter plane 
constructed from the grid search.  We identified two sets of local solutions, in which one set 
has a mass ratio between the lens components of $\log q\sim -2.0$ (Local A), and the other set 
has a mass ratio of $\log q \sim -4.1$ (Local B).  For each set, there exist a pair of solutions, 
designated as "inner" and "outer" solutions, and thus we identified four local solutions in total: 
"A$_{\rm in}$", "A$_{\rm out}$", "B$_{\rm in}$", and "B$_{\rm out}$". Below, we explain the choice 
of the notations "inner" and "outer" used to designate the solutions. In Figure~\ref{fig:three}, 
we mark the individual local solutions in the $\Delta\chi^2$ map.

In Table~\ref{table:one}, we list the lensing parameters of the individual local solutions 
together with $\chi^2$ values of the fit and degree of freedom. For each solution, the lensing 
parameters are refined by permitting them to vary from the initial values found from the 
first-round of modeling using the grid approach. The binary parameters of the solutions 
A$_{\rm in}$ and A$_{\rm out}$ are $(s, q)_{\rm in}\sim (1.37, 9.7\times 10^{-3})$ and 
$(s,~q)_{\rm out}\sim (0.77, 10.0\times 10^{-3})$, respectively, and those of the solutions 
B$_{\rm in}$ and B$_{\rm out}$ are $(s, q)_{\rm in}\sim (1.054, 6.6\times 10^{-5})$ and 
$(s, q)_{\rm out}\sim (0.997, 6.2\times 10^{-5})$, respectively. Across all solutions, the 
mass ratios between the lens components remain below $10^{-2}$, affirming that the companion 
to the lens is a planetary-mass object.  The local solutions B$_{\rm in}$ and B$_{\rm out}$ 
are preferred over their counterparts A$_{\rm in}$ and A$_{\rm out}$ by $\Delta\chi^2 = 12.9$ 
and 6.1, respectively.

\begin{figure}[t]
\includegraphics[width=\columnwidth]{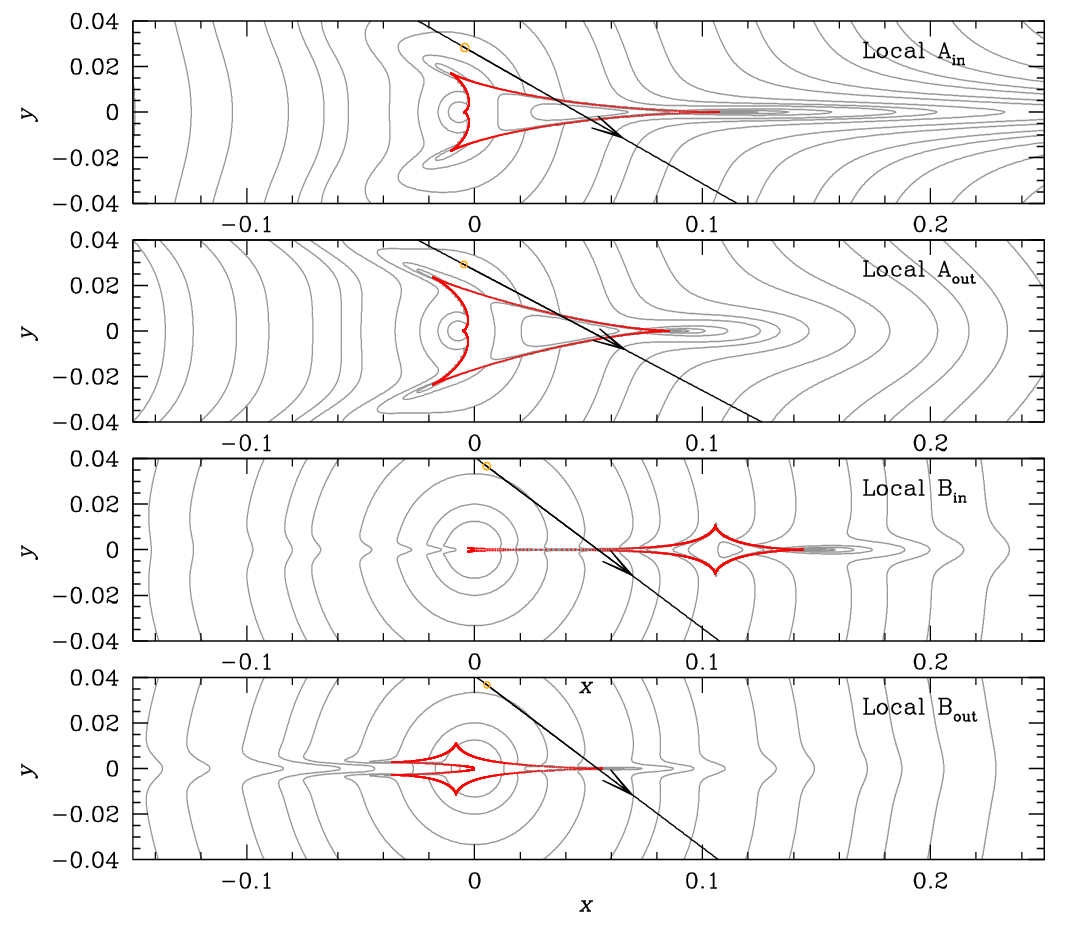}
\caption{
Lens-system configurations of the four local solutions of KMT-2023-BLG-0416. In each panel, 
the red cuspy figure represents the caustic, the arrowed line is the source trajectory, 
and the grey curves surrounding the caustic represent the equi-magnification contours. 
}
\label{fig:five}
\end{figure}

In Figure~\ref{fig:four}, we present the model curves of the A$_{\rm out}$ (dotted curve) 
and B$_{\rm out}$ (solid curve) solutions in the region around the anomaly. We note that 
the model curves of the A$_{\rm in}$ and B$_{\rm in}$ solutions are similar to those of the 
corresponding outer solutions A$_{\rm out}$ and B$_{\rm out}$.  Although the solution B 
provides a slightly better fit to the anomaly than the solution A, both solutions approximately 
describe the anomaly feature.  However, their model curves diverge significantly after the last 
data point of the anomaly within the time range $54.9\lesssim {\rm HJD}^\prime \lesssim 55.8$. 
This discrepancy suggests that the apparent degeneracy between solutions A and B is likely due 
to the incomplete coverage of the anomaly rather than a fundamental physical degeneracy.

Figure~\ref{fig:five} shows the lens-system configurations of the four degenerate solutions. 
From the comparison of the configurations, we find that the similarities between the model 
curves of the pair of the A$_{\rm in}$--A$_{\rm out}$ solutions and the pair of the 
B$_{\rm in}$--B$_{\rm out}$ solutions are caused by the "inner-outer" degeneracy. This 
degeneracy was originally proposed by \citet{Gaudi1997} to point out the similarity between 
the planetary signals produced by the source passage through the near (inner) and far (outer) 
sides of the planetary caustic.  Later \citet{Yee2021} found that the degeneracy can 
be extended to planetary signals induced by central and resonant caustics. \citet{Hwang2022} 
and \citet{Gould2022a} derived an analytic relation between the binary separations of the 
inner and outer solutions, $s_{\rm in}$ and $s_{\rm out}$. The relation is expressed as
\begin{equation}
\sqrt{s_{\rm in}\times s_{\rm sout}} = 
{  \sqrt{u_{\rm anom}^2 + 4}\pm u_{\rm anom}
\over 2},
\label{eq1}
\end{equation}
where $u_{\rm anom} = (\tau_{\rm anom}^2 + u_0^2)^{1/2}$, $\tau_{\rm anom} = (t_{\rm anom}-t_0)
/\te$, $t_{\rm anom}$ denotes the time of the planet-induced anomaly, and the "$+$" and "$-$" 
signs apply to the anomalies exhibiting positive and negative deviations, resulting from the major 
and minor image perturbations, respectively.  With the values of $(t_0, u_0, \te, t_{\rm anom}) 
\sim  (54.2, 0.023, 28.0, 55.4)$ for the pair of the A$_{\rm in}$--A$_{\rm out}$ solutions, we 
find that $(\sqrt{u_{\rm anom}^2 + 4}+u_{\rm anom})/2\sim  1.032$, which matches well the geometric 
mean of the binary separations of $\sqrt{s_{\rm in}\times s_{\rm sout}}\sim 1.029$.  For the pair 
of the B$_{\rm in}$--B$_{\rm out}$ solutions with $(t_0, u_0, \te, t_{\rm anom}) \sim (53.9, 0.033, 
24, 54.9)$, we find that $(\sqrt{u_{\rm anom}^2 + 4}+u_{\rm anom})/2\sim  1.027$, which also matches 
very well the geometric mean $\sqrt{s_{\rm in}\times s_{\rm sout}}\sim 1.025$.  The fact that the 
binary separations of the pair of solutions with similar model curves well follow the relation in 
Eq.~(\ref{eq1}) indicates that the model curves are subject to the inner--outer degeneracy. In 
Figure~\ref{fig:two}, we mark the geometric mean of $s_{\rm in}$ and $s_{\rm out}$ for the B 
solutions as a vertical dot-dashed line.

\begin{figure}[t]
\includegraphics[width=\columnwidth]{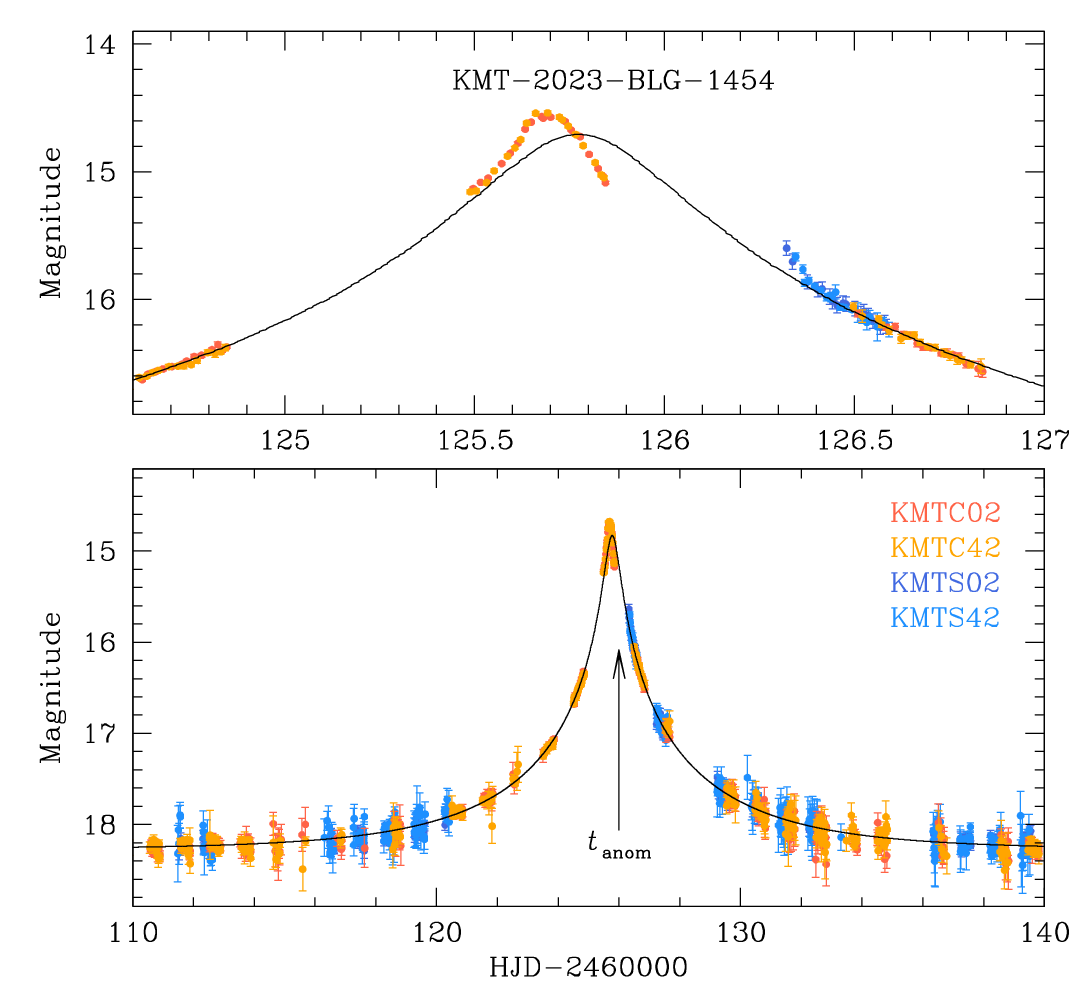}
\caption{
Light curve of KMT-2023-BLG-1454.  Notations are the same as those in Fig.~\ref{fig:two}. 
}
\label{fig:six}
\end{figure}

\begin{figure}[t]
\includegraphics[width=\columnwidth]{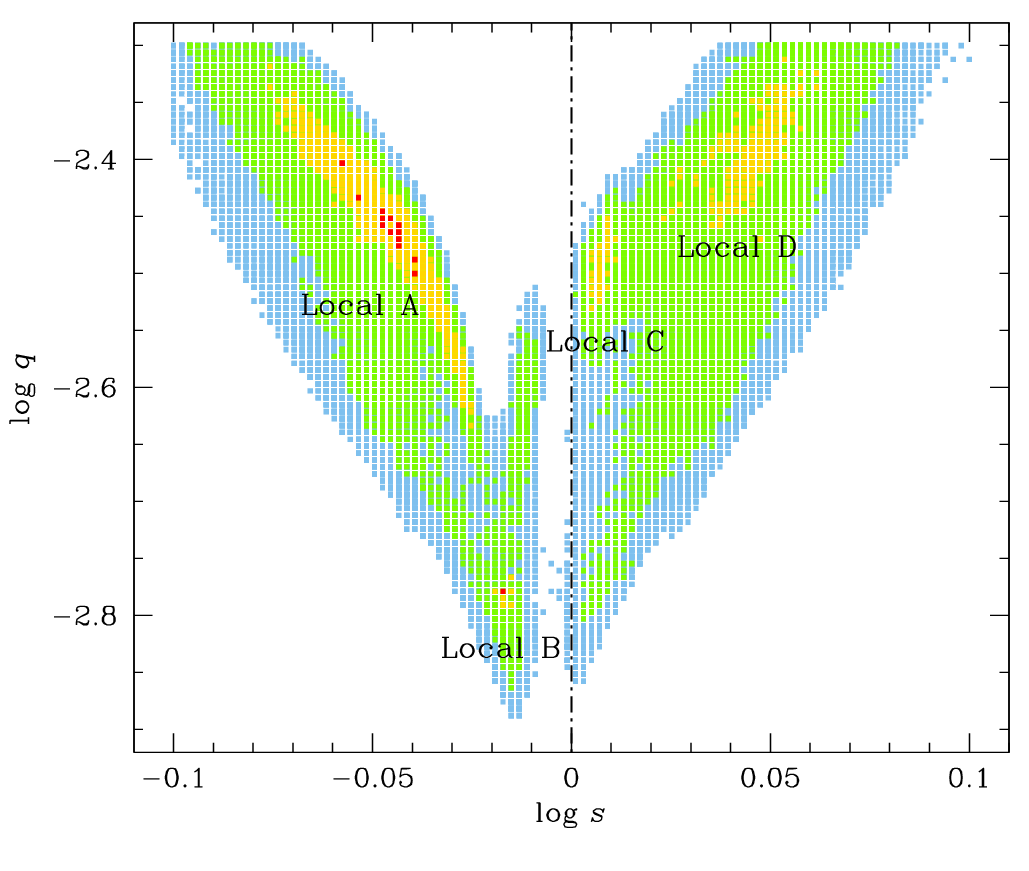}
\caption{
$\Delta\chi^2$ map in the $(\log~s, \log~q)$ parameter plane of KMT-2023-BLG-1454. The
notations and color coding of points are the same as those in Fig.~\ref{fig:two}. 
}
\label{fig:seven}
\end{figure}

Because it is known that a short-term positive deviation can also be produced by a faint companion 
to the source \citep{Gaudi1998}, we additionally checked for a possible binary-source origin of 
the anomaly. The model curve of the single-lens binary-source (1L2S) solution and the residual 
from the model are presented in Figure~\ref{fig:four}. We find that the 1L2S solution yields a 
model that is worse than the 2L1S model by $\Delta\chi^2 = 24.6$, which is most pronounced in 
the region just before the major positive anomaly.  As a result, we rule out the binary-source 
explanation for the anomaly.

\begin{table*}[t]
\footnotesize
\caption{Solutions of KMT-2023-BLG-1454}\label{table:two}
\begin{tabular}{lllll}
\hline\hline
\multicolumn{1}{c}{Parameter}                    &
\multicolumn{1}{c}{Local A}          &
\multicolumn{1}{c}{Local B}          &
\multicolumn{1}{c}{Local C}          &
\multicolumn{1}{c}{Local D}         \\
\hline
$\chi^2/{\rm dof}$            &  2924.3/2939              &  2934.7/2939             &  2934.2/2939             &   2936.3/2939             \\
$t_0$ (HJD$^\prime$)          &  $125.7451 \pm  0.0020$   &  $125.7642 \pm  0.0019$  &  $125.7514 \pm 0.0021$   &   $125.7464 \pm 0.0045$   \\
$u_0$ ($10^{-2}$)             &  $  4.085  \pm  0.060 $   &  $  3.849  \pm  0.055 $  &  $  4.219  \pm 0.063 $   &   $  4.110  \pm 0.074 $   \\
$\te$ (days)                  &  $  6.438  \pm  0.077 $   &  $  6.401  \pm  0.074 $  &  $  6.448  \pm 0.077 $   &   $  6.440  \pm 0.079 $   \\
$s$                           &  $  0.900  \pm  0.014 $   &  $  0.960  \pm  0.003 $  &  $  1.016  \pm 0.002 $   &   $  1.120  \pm 0.016 $   \\
$q$ ($10^{-3}$)               &  $  3.53   \pm  0.27  $   &  $  1.67   \pm  0.06  $  &  $  3.11   \pm 0.09  $   &   $  4.32   \pm 0.34  $   \\
$\alpha$ (rad)                &  $  0.966  \pm  0.011 $   &  $  1.152  \pm  0.017 $  &  $  0.958  \pm 0.010 $   &   $  0.964  \pm 0.035 $   \\
$\rho$ ($10^{-3}$)            &  $  21.49  \pm  0.44  $   &  $  21.83  \pm  0.36  $  &  $  21.93  \pm 0.32  $   &   $  20.79  \pm 0.42  $   \\
\hline                                                 
\end{tabular}
\end{table*}

\subsection{KMT-2023-BLG-1454}\label{sec:three-two}

The lensing event KMT-2023-BLG-1454 occurred on a star lying at the equatorial and Galactic 
coordinates of $({\rm RA}, {\rm DEC}) = $ (17:50:28.82, $-$29:37:21.90) and $(l, b) = 
(-0^\circ\hskip-2pt .040, -1^\circ\hskip-2pt .262)$, respectively. The source has a baseline 
magnitude $I_{\rm base} = 18.23$, and the extinction toward the field is $A_I = 3.09$. The 
event was found by the KMTNet group on 2023 June 29 (${\rm HJD}^\prime =124$).  One day after 
the event detection, the event reached its peak with a magnification of $A_{\rm peak}\sim 34$. 
The duration of the event is relatively short, and the source flux returned to its baseline 
about a week after reaching the peak.

\begin{figure}[t]
\includegraphics[width=\columnwidth]{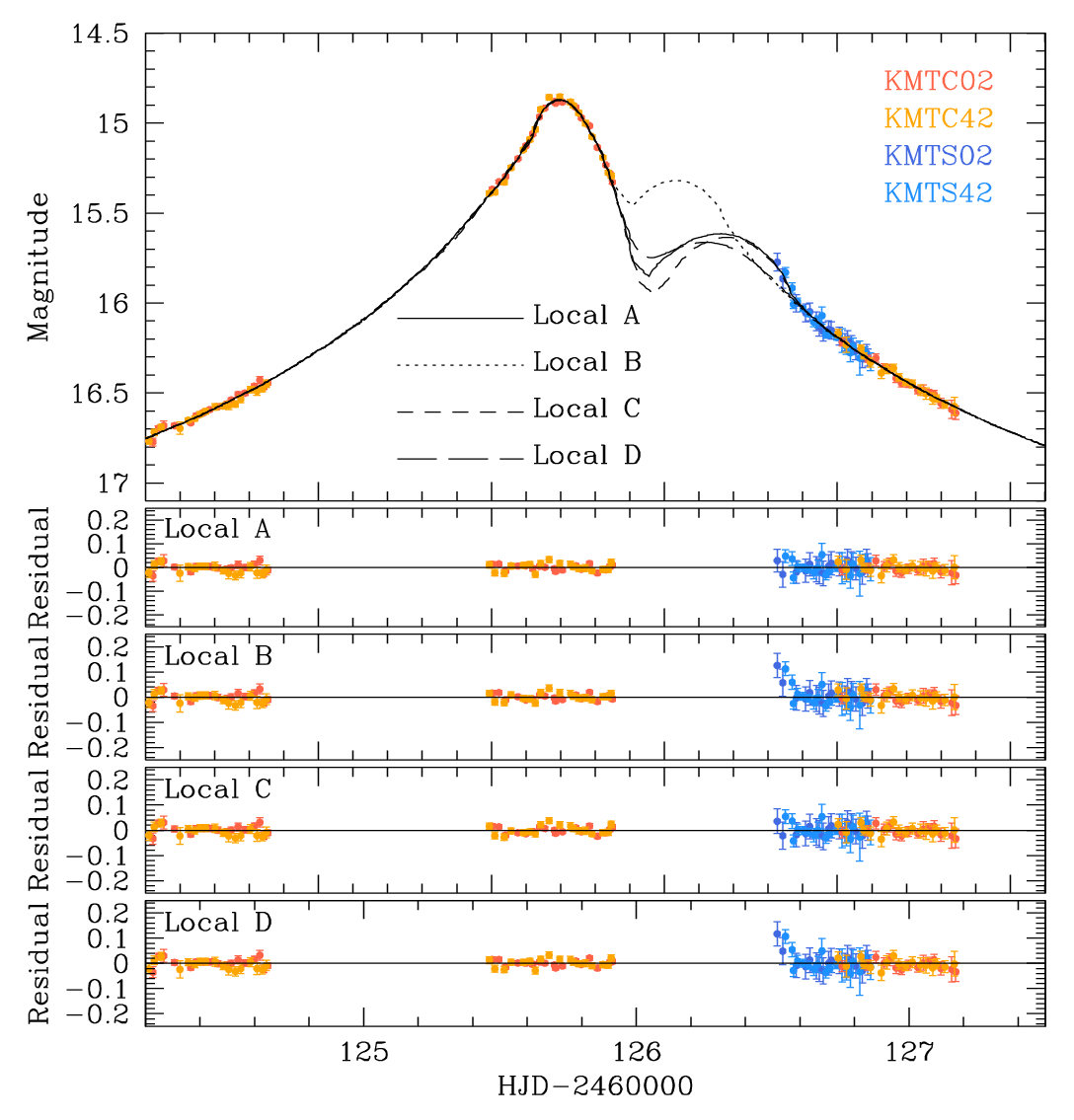}
\caption{
Model curves of the three local solutions (A, B, and C) of KMT-2023-BLG-1454 in the region 
of the anomaly. The three lower panels show the residuals from the individual solutions. 
}
\label{fig:eight}
\end{figure}

\begin{figure}[t]
\includegraphics[width=\columnwidth]{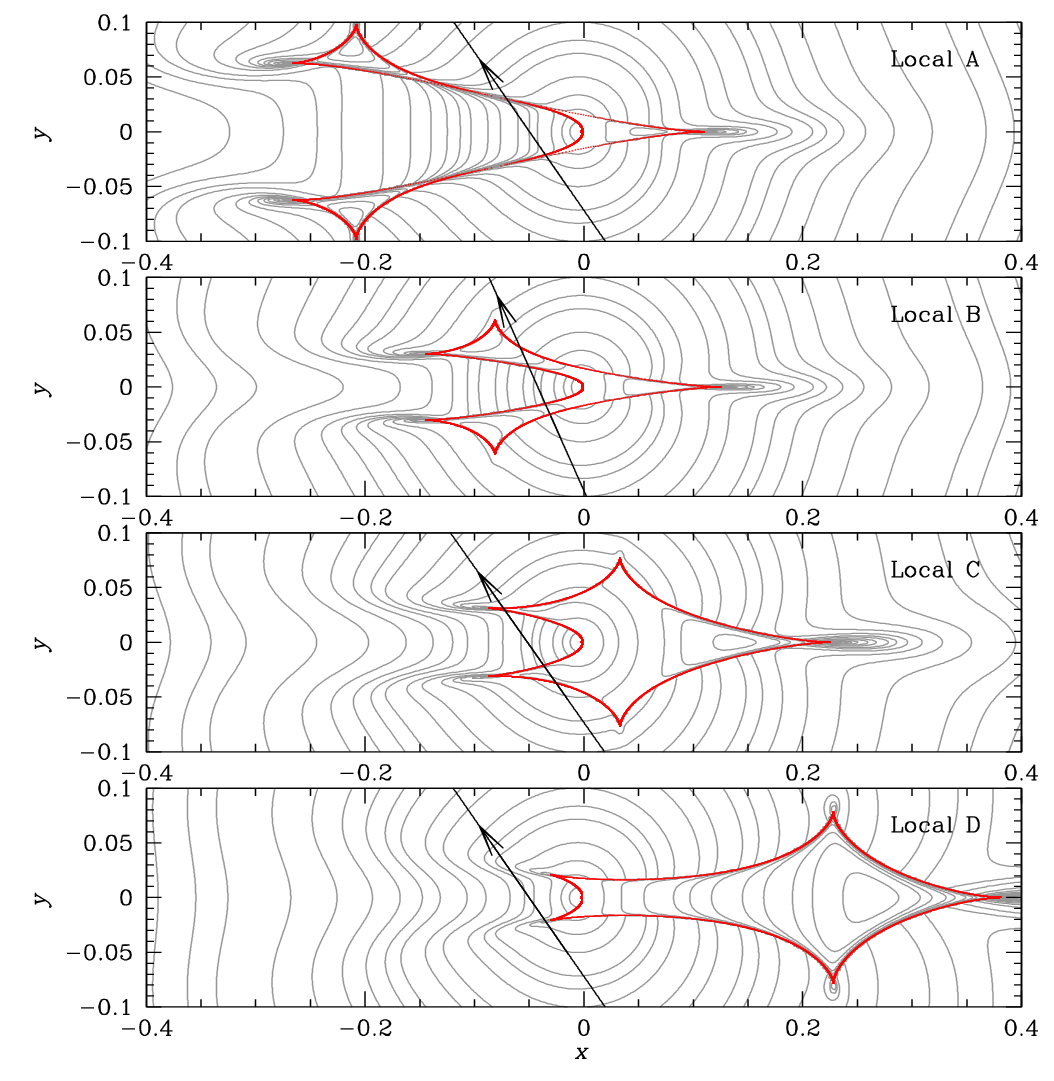}
\caption{
Lens-system configurations of the three degenerate solutions (inner, outer, and
intermediate solutions) of KMT-2023-BLG-1454. 
}
\label{fig:nine}
\end{figure}

In Figure~\ref{fig:six}, we present the light curve of KMT-2023-BLG-1454. The source of the 
event lies in the overlapping region of the two KMTNet prime fields BLG02 and BLG42, for which 
observations were conducted at a cadence of 0.5 hours for each field and 0.25 hours for the 
overlapping area of the two fields. In the analysis, we do not include the KMTA dataset not 
only because the peak region, near which an anomaly occurred, was not covered by these data, 
but also because the uncertainties in these data are large. From inspection of the light curve, we 
found that an anomaly occurred near the peak, as shown in the upper panel of Figure~\ref{fig:six}. 
The anomaly displays two separate features, in which the feature centered at ${\rm HJD}^\prime 
\sim 125.7$ exhibits both rising and falling deviations with respect to a 1L1S model, and the 
other feature centered at ${\rm HJD}^\prime \sim 126.5$ shows only a falling deviation. The 
structure of the anomaly could not be fully delineated due to the absence of the KMTA dataset, 
which could have spanned the gap between the KMTC and KMTS data if it were not for the cloud 
cover at the Australian site.  While a companion to a source can produced anomalies with only 
a single anomaly feature, a companion to a lens can produced anomalies with multiple features.  
Therefore, the anomaly is likely to be produced by a lens companion.

Considering the characteristics of the anomaly, we conducted a 2L1S modeling of the lensing 
light curve. The $\Delta\chi^2$ map on the $(\log s, \log q)$ parameter plane constructed 
from the grid search is shown in Figure~\ref{fig:seven}. The map shows  four distinct local 
solutions, with binary parameters $(\log s, \log q)\sim (-0.045, -2.45)$ (local A), $\sim 
(-0.01, -2.78)$ (local B), $\sim (0.007, -2.50)$ (local C), and $\sim (0.05, -2.36)$ (local D).  
Although the mass ratios of the local solutions exhibit slight differences, they consistently 
represent planet-to-star mass ratios across all solutions.  In Table~\ref{table:two}, we list 
the refined lensing parameters of the individual local solutions together with the $\chi^2$ 
values of the fits.  It is found that the local solution A is preferred over the other solutions 
by $\Delta\chi^2 \geq 9.9$.  The time scale of the event, $\te \sim 6.4$~day, is short.  To be 
discussed below, both anomaly features centered at ${\rm HJD}^\prime\sim 125.7$ and $\sim 126.5$ 
were produced by the source crossings over a caustic, and thus the normalized source radius, 
$\rho \sim 21.5 \times 10^{-3}$, is measured.

In Figure~\ref{fig:eight}, we present the model curves of the four local solutions together 
with the residuals from the models in the region around the anomaly. Figure~\ref{fig:nine} 
shows the configurations of the lens systems corresponding to the individual local solutions. 
For all solutions, the caustic displays a resonant form, in which the planetary caustics are 
connected with the central caustics.  It is found that the solutions A and D are the pair of 
solutions arising from the inner-outer degeneracy. According to the configurations, the source 
passed the inner region between the central and planetary caustic in the case of the A solution 
(inner solution), while the source passed the outer region of the planetary caustic in the case 
of the D solution (outer solution).  With the parameters $(t_0, u_0, \te, t_{\rm anom}) \sim  
(125.75, 0.041, 6.44, 126.0)$ for this pair of solutions, we find that $(\sqrt{u_{\rm anom}^2 
+ 4}- u_{\rm anom})/2\sim 1.03$, which matches very well the value $\sqrt{s_{\rm in}\times 
s_{\rm sout}} \sim 1.00$.  On the other hand, the degeneracies among these solutions and the 
other solutions are accidental arising  due to the absence of data in the region between the 
KMTC and KMTS datasets.  This is evidence by the substantial differences between the model 
light curves in the region that is not covered by data.

\begin{figure}[t]
\includegraphics[width=\columnwidth]{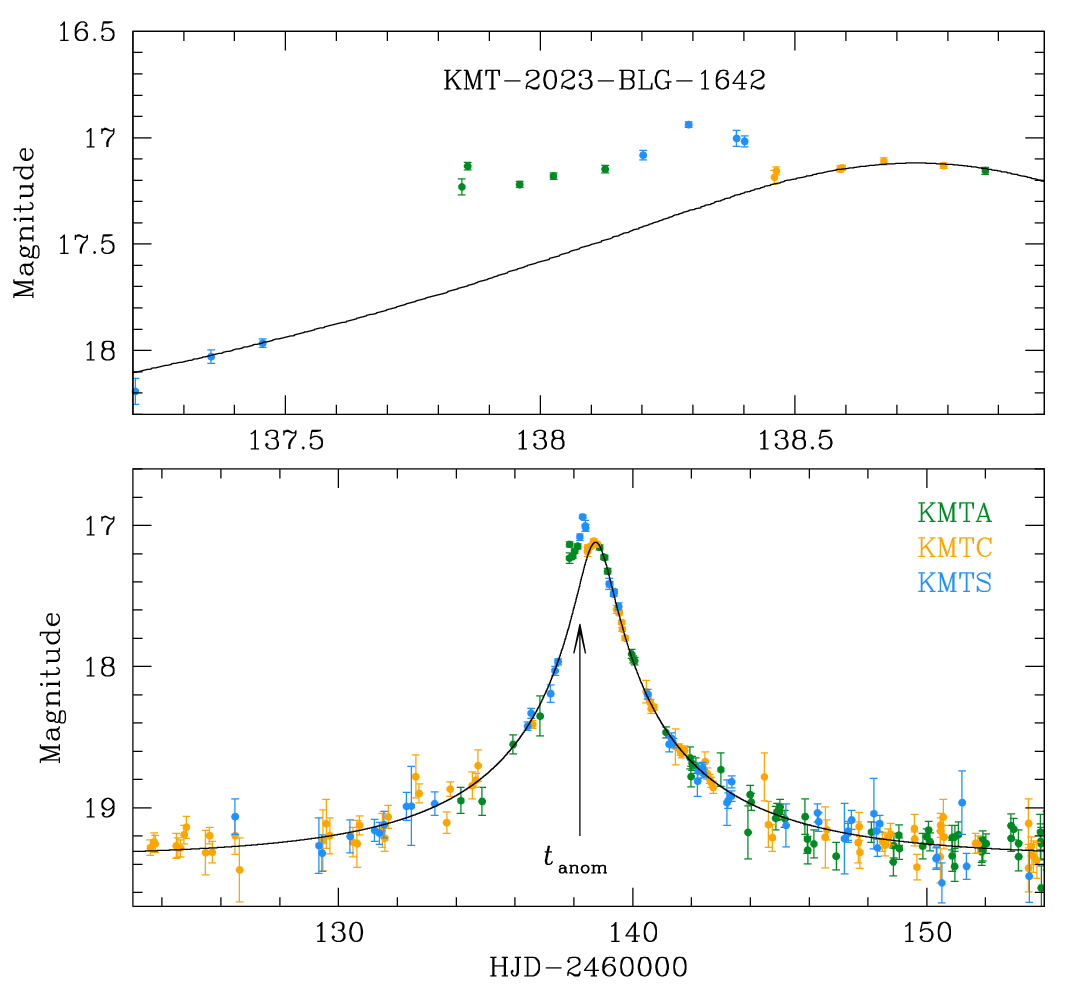}
\caption{
Lensing light curve of KMT-2023-BLG-1642. 
}
\label{fig:ten}
\end{figure}

\begin{figure}[t]
\includegraphics[width=\columnwidth]{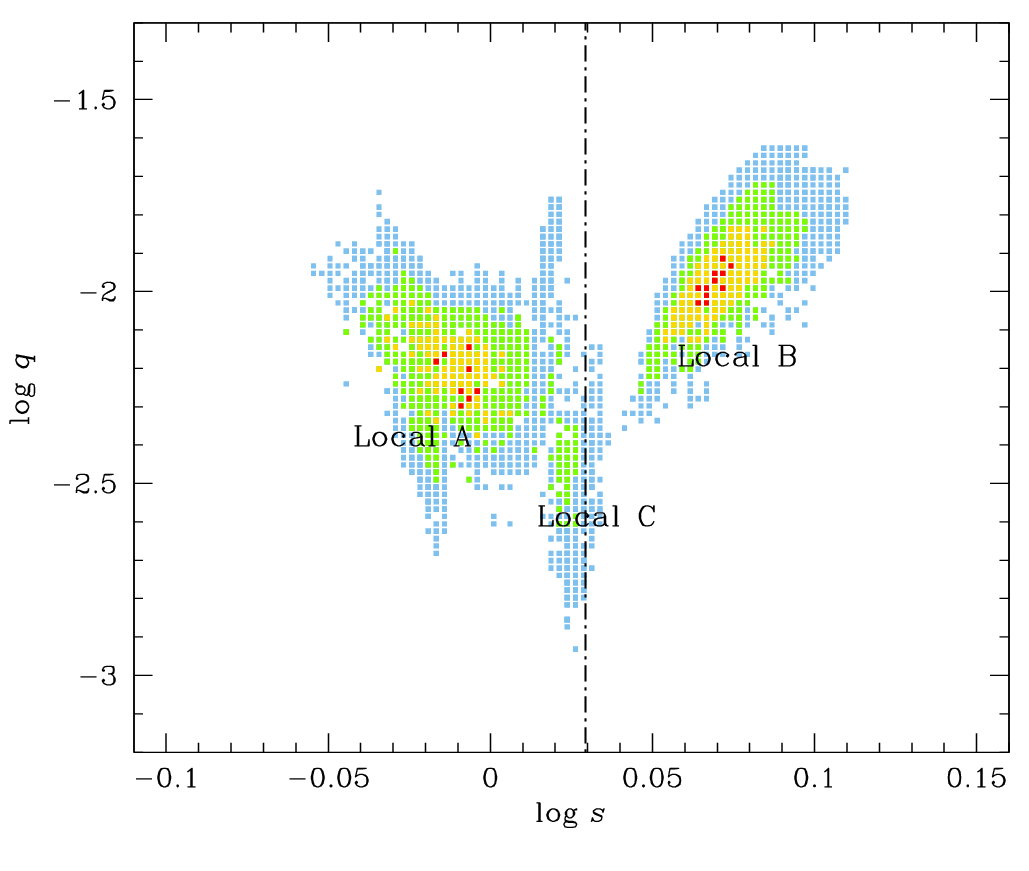}
\caption{
$\Delta\chi^2$ map in the $(\log s, \log q)$ parameter plane of KMT-2023-BLG-1642. The color 
coding is set to represent points with $\leq 1n\sigma$ (red), $\leq 2n\sigma$ (yellow), $\leq 
3n\sigma$ (green), and $\leq 4n\sigma$ (cyan), where $n=3$. 
}
\label{fig:eleven}
\end{figure}

\begin{table*}[t]
\footnotesize
\caption{Solutions of KMT-2023-BLG-1642}\label{table:three}
\begin{tabular}{lllll}
\hline\hline
\multicolumn{1}{c}{Parameter}                    &
\multicolumn{1}{c}{Local A}          &
\multicolumn{1}{c}{Local B}          &
\multicolumn{1}{c}{Local C}         \\
\hline
$\chi^2/{\rm dof}$            &   680.1/672                &   693.1/672              &   736.7/672                \\
$t_0$ (HJD$^\prime$)          &  $138.6337 \pm 0.0063$     &  $138.6268 \pm 0.0068$   &  $138.6444 \pm 0.0084$     \\
$u_0$ ($10^{-2}$)             &  $7.34 \pm 0.32        $   &  $7.33 \pm 0.42        $ &  $10.21 \pm 0.49       $   \\
$\te$ (days)                  &  $7.70 \pm 0.25        $   &  $7.52 \pm 0.35        $ &  $6.30 \pm 0.22        $   \\
$s$                           &  $0.9813 \pm 0.0037    $   &  $1.1666 \pm 0.0080    $ &  $1.0477 \pm 0.0025    $   \\
$q$ ($10^{-3}$)               &  $5.96 \pm 0.32        $   &  $10.44 \pm 0.83       $ &  $3.50 \pm 0.32        $   \\
$\alpha$ (rad)                &  $5.696 \pm 0.028      $   &  $5.464 \pm 0.017      $ &  $5.720 \pm 0.012      $   \\
$\rho$ ($10^{-3}$)            &  $5.78 \pm 0.60        $   &  $9.27 \pm 1.3         $ &  $5.81 \pm 1.12        $   \\
\hline                                                 
\end{tabular}
\end{table*}

\subsection{KMT-2023-BLG-1642}\label{sec:three-three}

\begin{figure}[t]
\includegraphics[width=\columnwidth]{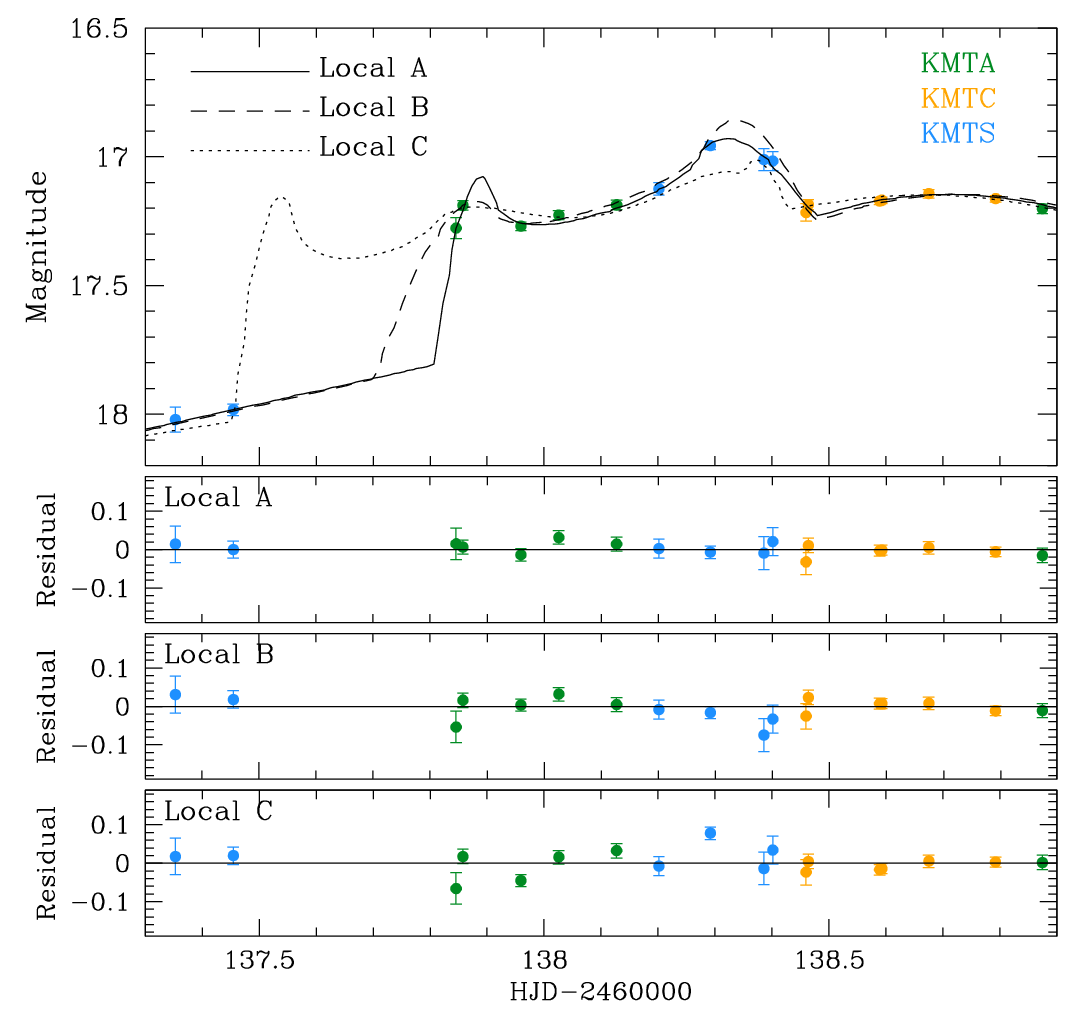}
\caption{
Model curves and residuals from the three local solutions of KMT-2023-BLG-1642 in the region 
of the anomaly.
}
\label{fig:twelve}
\end{figure}

\begin{figure}[t]
\includegraphics[width=\columnwidth]{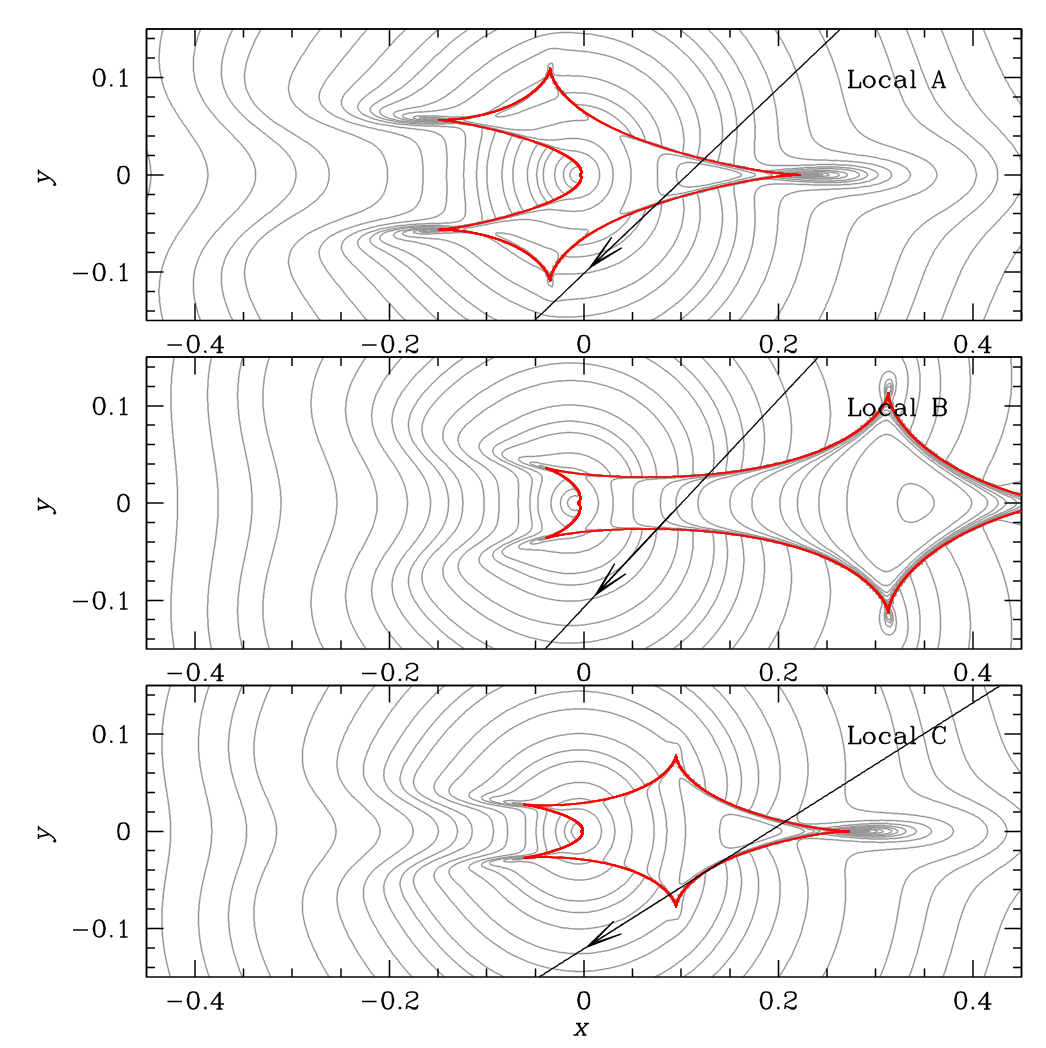}
\caption{
Lens-system configurations of the three local solutions of KMT-2023-BLG-1642. 
}
\label{fig:thirteen}
\end{figure}

\begin{figure*}[t]
\centering
\includegraphics[width=12.0cm]{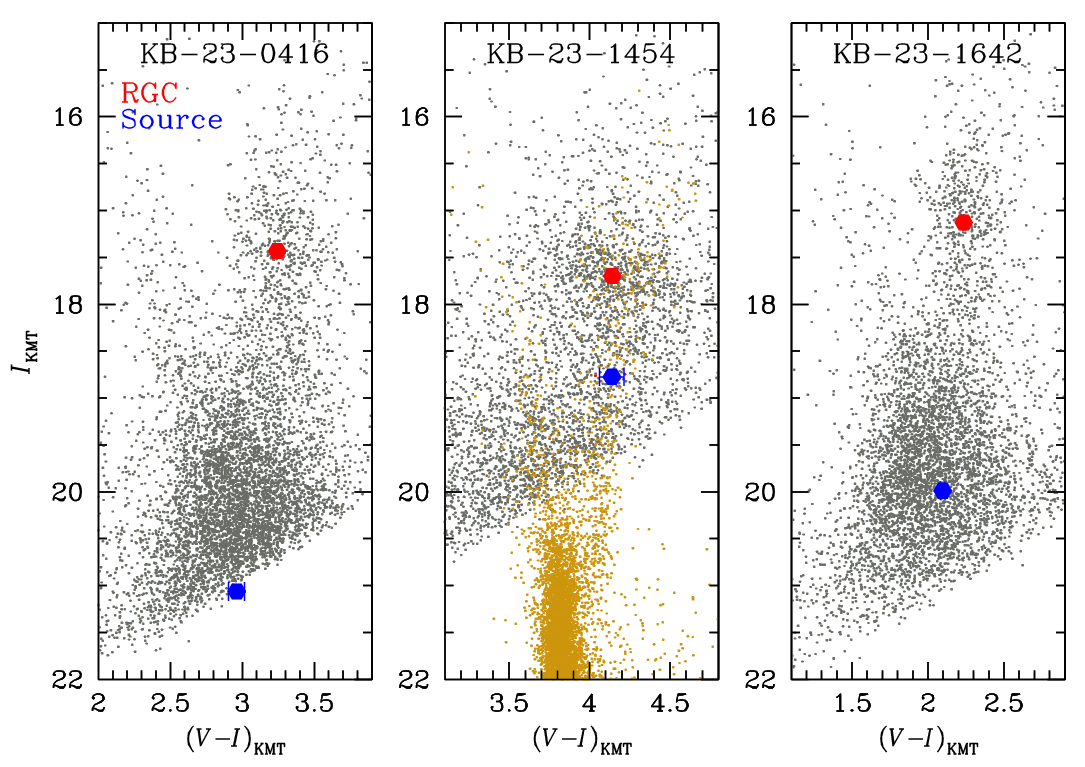}
\caption{
Locations of source stars (marked by blue dots) with respect to the red giant clump (RGC) 
centroids (marked by red dots) for the lensing events KMT-2023-BLG-0416, KMT-2023-BLG-1454, 
and KMT-2023-BLG-1642 in the instrumental color-magnitude diagrams of stars lying near the 
source stars of the events.  For KMT-2023-BLG-1454, the CMD is constructed by aligning the 
CMD established using KMTC data (grey dots) and that of stars in the Baade's window (brown 
dots) observed with the use of the Hubble Space Telescope.
}
\label{fig:fourteen}
\end{figure*}

The source of the lensing event KMT-2023-BLG-1642 lies at the equatorial coordinates $({\rm RA}, 
{\rm DEC}) = $ (17:25:27.84, $-29$:03:30.49), which correspond to the Galactic coordinates $(l, b) 
= (-2^\circ\hskip-2pt .482, 3^\circ\hskip-2pt .645)$.  The source brightness at the baseline is
$I_{\rm base}=19.34$, and the extinction toward the field is $A_I=1.87$. The KMTNet group alerted 
the event at UT 02:28 on 2023 July 14, which corresponds to ${\rm HJD}^\prime =140.1$. The event 
reached its peak at ${\rm HJD}^\prime =138.6$ with a magnification $A_{\rm peak}\sim 12.5$.  The 
duration of the event is relatively short with an event time scale of $\te \sim 7$~days.

Figure~\ref{fig:ten} displays the lensing light curve of KMT-2023-BLG-1642. Detailed inspection 
of the light curve revealed that there was an anomaly that occurred just before the event reached 
its peak. The anomaly lasted for about 1~day during $137.8\lesssim {\rm HJD}^\prime\lesssim 138.7$. 
From the anomaly structure, which is characterized by two bumps centered at ${\rm HJD}^\prime \sim 
137.85$ and $\sim 138.80$ and the U-shaped trough between the bumps, it is likely that the anomaly 
was produced by the caustic crossings of the source through a tiny caustic induced by a lens 
companion.  The  early part of the first bump was not be covered because the KMTC site was clouded 
out on the night of the anomaly.

Figure~\ref{fig:eleven} shows the $\Delta\chi^2$ map on the $\log s$--$\log q$ parameter plane 
constructed from the grid searches of the parameters. We identified three locals lying at 
$(\log s, \log q)\sim (-0.01, -2.22)$ (local A), $\sim (0.07, -1.98)$ (local B), and $\sim (0.02, 
-2.46)$ (local C). The full lensing parameters of the individual local solutions determined after 
the refinement are listed in Table~\ref{table:three}.  The estimated mass ratios between the lens 
components are $q \lesssim 10^{-2}$ regardless of the solutions, indicating that the anomaly was 
produced by a companion having a planetary mass.  In Figure~\ref{fig:twelve}, we present the model 
curves of the three local solutions and residuals from the models. From the comparison of the 
goodness of the fits, it is found that the solution~A is favored over the solutions~B and C by 
$\Delta\chi^2 =13.0$ and 56.3, respectively. Considering the $\chi^2$ differences, the solution~C 
is ruled out, but the solution~B cannot be completely excluded. For the solutions~A and B, the 
value $(\sqrt{u_{\rm anom}^2 + 4}+u_{\rm anom})/2\sim 1.05$ estimated from the lensing parameters 
approximately matches the geometric mean of the binary separations, $\sqrt{s_{\rm in}\times 
s_{\rm sout}}\sim 1.07$, of the two solutions, and thus the similarity between the model curves 
is caused by the inner--outer degeneracy. In Figure~\ref{fig:eleven}, we mark the geometric mean of 
$s_{\rm in}$ and $s_{\rm out}$ as a vertical dot-dashed line. The major difference between the 
model curves of the two solutions appears in the rising part of the first bump, but the incomplete 
coverage of the region makes it difficult to clearly lift the degeneracy between the solutions.

Figure~\ref{fig:thirteen} shows the lens-system configurations corresponding to the three local 
solutions. For all solutions, the caustics have a resonant form with connected planetary and 
central caustics. In the cases of the two solutions A and B, for which the lensing parameters 
except for the planet separation $s$ are similar to each other, the source passes the outer and 
inner regions of the planetary caustic, respectively, indicating that the similarity between the 
model curves of the two solutions is caused by the inner--outer degeneracy.  The lensing parameters 
$(u_0, \te, \rho)$ of the solution~C are substantially different from those of the other solutions, 
and this indicates that the degeneracy between this and the other solutions is accidental, mostly 
because of the incomplete coverage of the anomaly.

\section{Source stars and angular Einstein radii}\label{sec:four}

In this section, we specify the source stars of the individual lensing events.  Specifying 
the source of a lensing event is important not only to fully characterize the event but also 
to determine the angular Einstein radius, which is estimated by the relation
\begin{equation}
\thetae = {\theta_* \over \rho}, 
\label{eq2}
\end{equation}
where the normalized source radius is measured from the light curve modeling, and the angular
source radius $\theta_*$ can be deduced from the type of the source.

\begin{table*}[t]
\footnotesize
\caption{Source parameters}\label{table:four}
\begin{tabular}{lllll}
\hline\hline
\multicolumn{1}{c}{Parameter}        &
\multicolumn{1}{c}{KMT-2023-BLG-0416}          &
\multicolumn{1}{c}{KMT-2023-BLG-1454}          &
\multicolumn{1}{c}{KMT-2023-BLG-1642}         \\
\hline
 $(V-I, I)$                &  $(2.960 \pm 0.056, 21.062 \pm 0.004)$                &  $(4.139 \pm 0.076, 18.773 \pm 0.028)$  &  $(2.097 \pm 0.014, 19.989 \pm 0.003)$  \\
 $(V-I, I)_{\rm RGC}$      &  $(3.242, 17.430)                    $                &  $(4.141, 17.700)                    $  &  $(2.235 \pm 0.040, 17.128 \pm 0.020)$  \\
 $(V-I, I)_{\rm RGC,0}$    &  $(1.060, 14.593)                    $                &  $(1.060, 14.445)                    $  &  $(1.060, 14.593)                    $  \\
 $(V-I, I)_0$              &  $(0.778 \pm 0.069, 18.225 \pm 0.020)$                &  $(1.058 \pm 0.076, 15.519 \pm 0.028)$  &  $(0.923 \pm 0.042, 17.454 \pm 0.020)$  \\
\hline                                                                                                                                                                 
\end{tabular}
\end{table*}

\begin{table*}[t]
\footnotesize
\caption{Einstein radii and relative lens-source proper motions}\label{table:five}
\begin{tabular}{lllll}
\hline\hline
\multicolumn{1}{c}{Parameter}                  &
\multicolumn{1}{c}{KMT-2023-BLG-0416}          &
\multicolumn{1}{c}{KMT-2023-BLG-1454}          &
\multicolumn{1}{c}{KMT-2023-BLG-1642}         \\
\hline
 $\theta_*$ (uas)          &  $0.770 \pm 0.075$                         &  $3.741 \pm 0.386$               &  $1.312 \pm 0.107$              \\
\hline
 $\thetae$ (mas)           &  $0.439 \pm 0.093$ (Local A$_{\rm in}$)    &  $0.174 \pm 0.018                $  &  $0.227 \pm 0.051$ (Local A) \\
                           &  $0.546 \pm 0.139$ (Local A$_{\rm out}$)   &  --                              &  $0.142 \pm 0.022$ (Local B)    \\
                           &  $0.497 \pm 0.116$ (Local B$_{\rm in}$)    &  --                              &  --                             \\
                           &  $0.585 \pm 0.158$ (Local B$_{\rm out}$)   &  --                              &  --                             \\
\hline
 $\mu$ (mas/yr)            &  $5.43 \pm 1.13$   (Local A$_{\rm in}$)    &  $9.87 \pm 1.05$                 &  $10.77 \pm 2.44$  (Local A)    \\
                           &  $7.35 \pm 1.86$   (Local A$_{\rm out}$)   &  --                              &  $6.88  \pm 1.07$ (Local B)     \\
                           &  $7.84 \pm 1.83$   (Local B$_{\rm in}$)    &  --                              &  --                             \\
                           &  $8.80 \pm 2.37$   (Local B$_{\rm out}$)   &  --                              &  --                             \\
\hline                                                                                                                                                                 
\end{tabular}
\end{table*}

The source type is determined by measuring the reddening- and extinction-corrected (de-reddened) 
color and magnitude using the \citet{Yoo2004a} routine. In the first step of this routine, we 
measured the instrumental color and magnitude $(V-I, I)$ of the source by regressing the $I$- 
and $V$-band datasets processed using the pyDIA photometry code with respect to the model lensing 
light curve, and then placed the source in the instrumental color-magnitude diagram (CMD) of stars 
lying near the source constructed with the use of the same pyDIA code.  In the second step, we 
calibrated the color and magnitude of the source using the centroid of the red giant clump (RGC) 
as a reference, that is, 
\begin{equation}
(V-I, I)_0 = (V-I, I)_{\rm RGC,0} + \Delta(V-I, I). 
\label{eq3}
\end{equation}
Here $(V-I, I)_0$ and $(V-I, I)_{\rm RGC,0}$ represent the de-reddened colors and magnitudes 
of the source and RGC centroid, respectively, and $\Delta(V-I, I)=(V-I, I)-(V-I, I)_{\rm RGC}$ 
represents the offsets in color and magnitude between the source and RGC centroid. The RGC 
centroid can be used as a reference for calibration because its de-reddened color and magnitude 
are known from \citet{Bensby2013} and \citet{Nataf2013}, respectively.

\begin{figure}[t]
\includegraphics[width=\columnwidth]{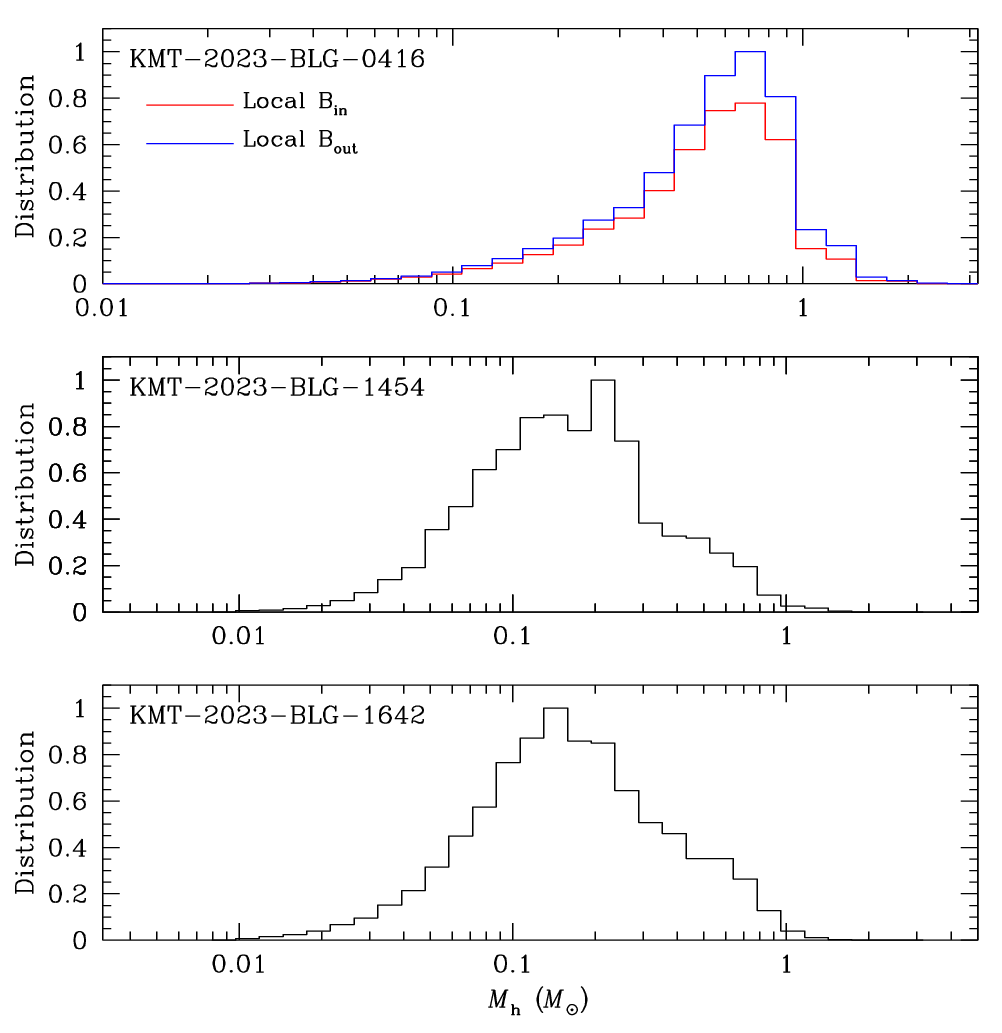}
\caption{
Bayesian posteriors for the host masses of the planetary systems KMT-2023-BLG-0416L, 
KMT-2023-BLG-1454L, and KMT-2023-BLG-1642L.  For KMT-2023-BLG-0416 and KMT-2023-BLG-1642 with 
varying $\thetae$ values depending on local solutions, multiple posteriors corresponding to 
the individual solutions are presented. For these events, the relative probability is scaled 
by applying a factor $\exp(-\Delta\chi^2/2)$, where $\Delta\chi^2$ denotes the difference in 
$\chi^2$ compared to the best-fit solution. The posterior of KMT-2023-BLG-1642 associated with 
the solution B is not prominently visible because of the very small scale factor. 
}
\label{fig:fifteen}
\end{figure}

In Figure~\ref{fig:fourteen}, we mark the position of the source star for each event and RGC 
centroid on the instrumental CMD.  For KMT-2023-BLG-1454, the $V$-band magnitude of the source 
could not be measured due to the combination of the limited number of the $V$-band data points 
during the lensing magnification and the low quality of the data caused by the relatively severe 
extinction toward the field.  In this case, we estimated the source color as the mean of colors 
of the stars lying in the giant or main-sequence branch of the combined ground+HST CMD with 
$I$-band magnitude offsets from the RGC centroid lying within the range of the measured value.  
The combined CMD was constructed by aligning the CMD established using KMTC data and that of 
stars in the Baade's window observed with the use of the Hubble Space Telescope \citep{Holtzman1998}.  
In Table~\ref{table:four}, we list the estimated values of $(V-I, I)$, $(V-I, I)_{\rm RGC}$, 
$(V-I, I)_{\rm RGC,0}$, and the finally determined de-reddened source colors and magnitudes, 
$(V-I, I)_0$, of the individual events.  According to the estimated colors and magnitudes, the 
source star is a late G-type turnoff star for KMT-2023-BLG-0416, a K-type giant for 
KMT-2023-BLG-1454, and an early K-type subgiant star for KMT-2023-BLG-1642.

For the estimation of the angular Einstein radius from the relation in Eq.~(\ref{eq2}), 
we initially converted the $V-I$ color into $V-K$ color using the \citet{Bessell1988} relation, 
and subsequently determined the angular source radius by applying the \citet{Kervella2004} 
relationship between $V-K$ color and $\theta_*$.  With the estimated value of $\thetae$, the 
relative lens-source proper motion is computed using the measured event time scale as 
\begin{equation}
\mu = {\thetae\over \te}.
\label{eq4}
\end{equation}
In Table~\ref{table:five}, we list the estimated values of $\theta_*$, $\thetae$, and $\mu$ 
for the individual lensing events.  For the events KMT-2023-BLG-0416 and KMT-2023-BLG-1642, 
the $\rho$ value varies substantially over the local solutions, and thus we estimate $\thetae$ 
and $\mu$ values corresponding to the individual local solutions.

\begin{figure}[t]
\includegraphics[width=\columnwidth]{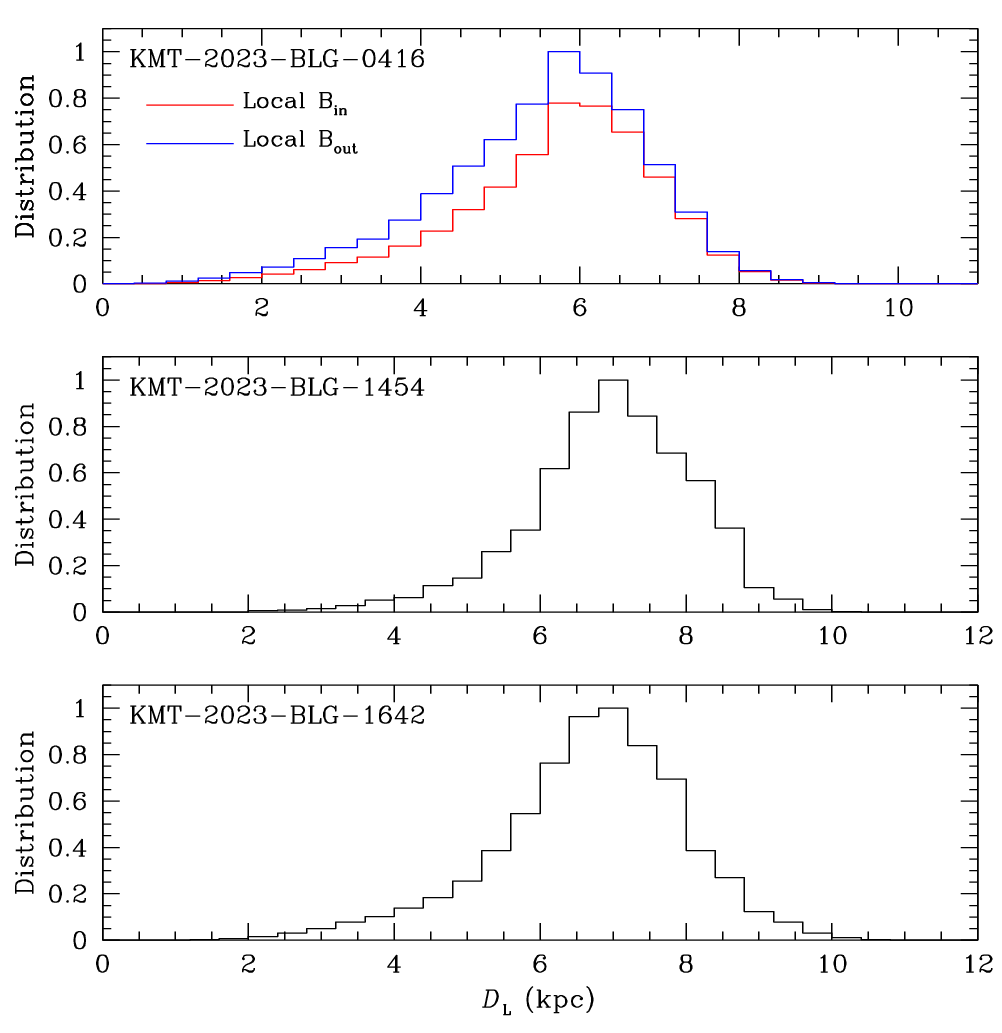}
\caption{
Bayesian posteriors for the distances to the planetary systems KMT-2023-BLG-0416L, KMT-2023-BLG-1454L, 
and KMT-2023-BLG-1642L.  Notations are the same as those in Fig.~\ref{fig:fifteen}.
}
\label{fig:sixteen}
\end{figure}

\begin{table*}[t]
\footnotesize
\caption{Physical lens parameters}\label{table:six}
\begin{tabular}{lllll}
\hline\hline
\multicolumn{1}{c}{Event}                           &
\multicolumn{1}{c}{$M_{\rm h}$ ($M_\odot$)}         &
\multicolumn{1}{c}{$M_{\rm p}$ ($M_{\rm J}$)}       &
\multicolumn{1}{c}{$\dl$ (kpc)}                     &
\multicolumn{1}{c}{$a_\perp$ (AU)}                  \\
\hline                                                           
KMT-2023-BLG-0416                                  &                         &                                 &                            &                         \\ 
Local A$_{\rm in}$   ($\chi^2=1379.3$)  & $0.61^{+0.30}_{-0.32}$  &  $6.15^{+3.03}_{-3.20}      $   &  $6.33^{+0.95}_{-1.35}$    &  $3.73^{+0.56}_{-0.80}$ \\  [0.7ex]
Local A$_{\rm out}$  ($\chi^2=1372.0$)  & $0.64^{+0.31}_{-0.32}$  &  $6.66^{+3.26}_{-3.38}      $   &  $5.96^{+1.04}_{-1.49}$    &  $2.32^{+0.40}_{-0.58}$ \\  [0.7ex]
Local B$_{\rm in}$   ($\chi^2=1366.4$)  & $0.61^{+0.31}_{-0.31}$  &  $0.042^{+0.021}_{-0.022}   $   &  $6.08^{+1.01}_{-1.39}$    &  $3.04^{+0.51}_{-0.70}$ \\  [0.7ex]
Local B$_{\rm out}$  ($\chi^2=1365.9$)  & $0.63^{+0.32}_{-0.32}$  &  $0.041^{+0.020}_{-0.021}   $   &  $5.90^{+1.07}_{-1.50}$    &  $2.96^{+0.54}_{-0.75}$ \\  [0.7ex]
\hline
KMT-2023-BLG-1454                                  &                         &                                 &                            &                         \\
Local A  ($\chi^2=2924.3$)              & $0.17^{+0.19}_{-0.09}$  &  $0.63^{+0.71}_{-0.34}$         &  $7.22^{+1.06}_{-1.10}$    &  $1.18^{+0.17}_{-0.18}$ \\  [0.7ex]
Local B  ($\chi^2=2934.7$)              & --                      &  $0.30^{+0.34}_{-0.16}$         &  --                        &  $1.26^{+0.19}_{-0.19}$ \\  [0.7ex]
Local C  ($\chi^2=2934.2$)              & --                      &  $0.56^{+0.63}_{-0.30}$         &  --                        &  $1.33^{+0.20}_{-0.20}$ \\  [0.7ex]
Local D  ($\chi^2=2936.3$)              & --                      &  $0.77^{+0.87}_{-0.42}$         &  --                        &  $1.47^{+0.22}_{-0.22}$ \\  [0.7ex]
\hline
KMT-2023-BLG-1642                       &                         &                                 &                            &              \\
Local A ($\chi^2=680.1$)                & $0.17^{+0.24}_{-0.09}$  &  $1.08^{+1.53}_{-0.58}$         &  $6.98^{+1.09}_{-1.34}$    &  $1.41^{+0.22}_{-0.27}$ \\  [0.7ex]
Local B ($\chi^2=693.1$)                & $0.13^{+0.17}_{-0.06}$  &  $1.28^{+1.83}_{-0.68}$         &  $7.44^{+1.09}_{-1.15}$    &  $1.32^{+0.19}_{-0.20}$ \\
\hline                                                                                                                                   
\end{tabular}
\end{table*}

\section{Physical lens parameters}\label{sec:five}

In this section, we estimate the physical lens parameters of the individual events. The physical 
parameters of the lens mass $M$ and distance $\dl$ are constrained from the lensing observables 
of $\te$, $\thetae$, and $\pie$. Here $\pie$ indicates the microlens parallax, which is related 
to the relative lens-source parallax $\pi_{\rm rel} = {\rm AU}(1/\dl - 1/\ds)$ and proper motion 
by 
\begin{equation}
\pivec_{\rm E} =\left( {\pi_{\rm rel}\over \thetae} \right)\left( { \muvec \over \mu}\right),
\label{eq5}
\end{equation}
where $\ds$ denotes the distance to the source. With the simultaneous measurements of all these 
observables, the mass and distance to the lens are uniquely determined by the \citet{Gould2000} 
relation
\begin{equation}
M = {\thetae \over \kappa\pie};\qquad
\dl = {{\rm AU} \over \pie\thetae + \pi_{\rm S}},
\label{eq6}
\end{equation}
where $\kappa = 4G/(c^2{\rm AU}) \simeq 8.14~{\rm mas}/M_\odot$ and $\pi_{\rm S} = {\rm AU}/\ds$ 
denotes the parallax of the source. Among these lensing observables, the values of the event 
time scale and Einstein radius were securely measured for all events. The value of the microlens 
parallax can be measured from the subtle deviations in the lensing light curve induced by the 
digression of the relative lens-source motion from rectilinear caused by the orbital motion of 
Earth around the sun \citep{Gould1992a}. For none of the events, the microlens parallax can be 
measured because either the photometric precision of data is low or the event time scale is short.  
Nevertheless, the lens parameters can still be constrained because the other observables of $\te$ 
and $\thetae$ provide constraints on the mass and distance by the relations
\begin{equation}
\te = {\thetae \over \mu };\qquad
\thetae = \left( \kappa M \pi_{\rm rel}\right)^{1/2}.
\label{eq7}
\end{equation}
Therefore, we estimate the physical lens parameters by conducting Bayesian analyses with the
constraints provided by $\te$ and $\thetae$ values of the individual events.

The Bayesian analysis was conducted according to the following procedure.  In the first 
step of this process, we carried out a Monte Carlo simulation to generate a large number 
of synthetic lensing events, with the use of the prior information on the location, velocity, 
and mass function of astronomical objects within the Galaxy.  In this simulation, we adopted 
the \citet{Jung2021} Galaxy model for the physical and dynamical distributions, and the 
\citet{Jung2022} model for the mass function of lens objects.  Regarding KMT-2023-BLG-1454, 
the Gaia data archive \citep{Gaia2018} contains information about the source, including its 
proper motion, but we have chosen not to include the Gaia proper motion value in our Monte Carlo 
simulation due to the substantial uncertainties associated with the source's faintness.  For each 
synthetic event produced from this simulation, we allocated the lens mass $M_i$, lens distance 
$D_{{\rm L},i}$, source distance $D_{{\rm S},i}$, and the lens-source transverse velocity 
$v_{\perp,i}$, and then computed the corresponding lens observables using the relations 
$t_{{\rm E},i}= D_{{\rm L},i}\theta_{{\rm E},i}/v_{\perp,i}$ and $\theta_{{\rm E},i} = (\kappa 
M_i \pi_{{\rm rel},i})^{1/2}$, and $\pi_{{\rm rel},i}={\rm AU}(1/D_{{\rm L},i}-1/D_{{\rm S},i})$.  
In the second step, we constructed posteriors of the lens mass and distance by assigning a 
weight to each synthetic event of 
\begin{equation}
w_i = \exp\left(-{{\chi^2}\over 2}\right);\qquad
\chi^2 = \left[ {(t_{{\rm E},i}-\te)^2 \over \sigma^2(\te)} \right] + 
\left[ {(\theta_{{\rm E},i}-\thetae)^2 \over \sigma^2(\thetae)} \right].
\label{eq8}
\end{equation}
Here $(\te, \thetae)$ denote the measured values of the lensing observables, and $[\sigma(\te), 
\sigma(\thetae)]$ represent their uncertainties.

In Figures~\ref{fig:fifteen} and \ref{fig:sixteen}, we present the posteriors of the masses of 
the host stars and distances to the individual planetary systems.  In the cases of KMT-2023-BLG-0416 
and KMT-2023-BLG-1642, for which local solutions yield substantially divergent $\thetae$ values, 
we constructed posteriors corresponding to the individual solutions.  To represent these posteriors, 
we apply a scaling based on $\exp(-\Delta\chi^2/2)$, where $\Delta\chi^2$ denotes the difference 
in $\chi^2$ compared to the best-fit solution.  We point out that the scaling factors are $\lesssim 
0.05$ for the A solutions of KMT-2023-BLG-0416, $\lesssim 0.007$ for the B, C, and D solutions of 
KMT-2023-BLG-1454, and $\lesssim 0.002$ for the A solution of KMT-2023-BLG-1642, and thus the 
posteriors associated with these solutions are not presented.  In Table~\ref{table:six}, we list 
the estimated values of the host mass $M_{\rm h}$, planet mass $M_{\rm p}=qM_{\rm h}$, distance 
$\dl$, and projected physical separation between the planet and host, $a_\perp=s\dl\thetae$.  For 
each physical parameter, we choose the median value of the posterior distribution as the representative 
value,  and set the 16\% and 84\% of the posterior distribution as the lower and upper limits, respectively.

According to the Bayesian estimates, The estimated host mass of KMT-2023-BLG-0416L corresponds to 
a main-sequence star of a late K spectral type, and those of KMT-2023-BLG-1454L and KMT-2023-BLG-1642L 
correspond to that of a mid-M dwarf.  In the case of the planetary system KMT-2023-BLG-0416L, the 
planet's mass exhibits significant variation across solutions, ranging from approximately 6 times 
the mass of Jupiter according to A solutions, but resembling the mass of Uranus according to B 
solutions.  In the case of the KMT-2023-BLG-1454L planetary system, the planet has a mass roughly 
half that of Jupiter, while in the case of the KMT-2023-BLG-1646L system, the planet exhibits a 
mass range of approximately 1.1 to 1.3 times that of Jupiter, categorizing both as giant planets.

\section{Summary and conclusion}\label{sec:six}

We presented the analyses of three lensing events KMT-2023-BLG-0416, KMT-2023-BLG-1454, and
KMT-2023-BLG-1642, for which partially-covered short-term signals were found in their light 
curves from the investigation of the 2023 season data obtained from high-cadence microlensing 
surveys.  Through these analyses, we identified that the signals in the analyzed lensing events 
were generated by planetary companions to the lenses.

Given the potential degeneracy caused by the partial coverage of the signals, we conducted a
thorough exploration of the lensing parameter space. From the analysis of KMT-2023-BLG-0416,
we have identified two distinct sets of solutions: one characterized by a mass ratio of 
approximately $q \sim 10^{-2}$, and the other with $q \sim 6.5 \times 10^{-5}$, with each set 
yielding a pair of solutions caused by the inner-outer degeneracy.  In the case of KMT-2023-BLG-1454, 
we have identified four distinct local solutions with mass ratios spanning in the range of $q \sim 
(1.7 - 4.3) \times 10^{-3}$.  Among these solutions, two displayed resemblances in their model 
curves due to the inner-outer degeneracy, while the other two solutions arose due to accidental 
degeneracy.  Regarding KMT-2023-BLG-1642, we have discerned two local solutions with mass ratios 
$q \sim (6 - 10) \times 10^{-3}$ arising from the inner-outer degeneracy.

We derived the physical lens parameters through Bayesian analyses, which incorporated constraints
from the measured lensing observables of the event time scale and Einstein radius in conjunction
with prior information regarding the density, velocity, and mass distribution of astronomical 
objects within our Galaxy.  For KMT-2023-BLG-0416L planetary system, the host mass is $\sim 0.6~M_\odot$ 
and the planet mass is $\sim (6.1 - 6.7)~M_{\rm J}$ according to one set of solutions and $\sim 0.04~M_{\rm J}$ 
according to the other set of solutions.  KMT-2023-BLG-1454Lb has a mass roughly half that of Jupiter, 
while KMT-2023-BLG-1646Lb has a mass in the range of between 1.1 to 1.3 times that of Jupiter, 
classifying them both as giant planets orbiting mid M-dwarf host stars with masses ranging from 
0.13 to 0.17 solar masses.

\begin{acknowledgements}
Work by C.H. was supported by the grants of National Research Foundation of Korea (2019R1A2C2085965). 
J.C.Y. acknowledges support from U.S. NSF Grant No. AST-2108414.
Y.S. acknowledges support from BSF Grant No. 2020740.
This research has made use of the KMTNet system operated by the Korea Astronomy and Space 
Science Institute (KASI) and the data were obtained at three host sites of CTIO in Chile, 
SAAO in South Africa, and SSO in Australia.
\end{acknowledgements}

\end{document}